\documentclass[aps,prb,reprint]{revtex4-2}

\usepackage{amsmath}
\usepackage{amssymb}
\usepackage{bm}
\usepackage[utf8]{inputenc}
\usepackage[english]{babel}
\usepackage{graphicx}
\usepackage{float}
\usepackage[section]{placeins}  
\usepackage{booktabs}
\usepackage{multirow}
\usepackage{makecell}
\usepackage{textcomp}
\usepackage{xfrac}
\setcitestyle{super} 

\begin{document}
\renewcommand{\abstractname}{}    
\renewcommand{\figurename}{Fig.}
\makeatletter  \renewcommand\@biblabel[1]{#1.}  \makeatother
\newcommand{\parameters}{\mathbf{w}}

\title{Autonomous artificial intelligence discovers mechanisms of molecular self-organization in virtual experiments}
\bibliographystyle{naturemag}

\author{Hendrik Jung}
\altaffiliation{These authors contributed equally.}
\affiliation{Department of Theoretical Biophysics, Max Planck Institute of Biophysics, 60438 Frankfurt am Main, Germany.}
\author{Roberto Covino}
\altaffiliation{These authors contributed equally.}
\affiliation{Frankfurt Institute for Advanced Studies, 60438 Frankfurt am Main, Germany.}
\author{A Arjun}
\author{Peter G. Bolhuis}
\affiliation{van 't Hoff Institute for Molecular Sciences, University of Amsterdam,  PO Box 94157, 1090 GD Amsterdam, The Netherlands.}
\author{Gerhard Hummer}
\email[Corresponding author:]{gerhard.hummer@biophys.mpg.de}
\affiliation{Department of Theoretical Biophysics, Max Planck Institute of Biophysics, 60438 Frankfurt am Main, Germany.}
\affiliation{Institute of Biophysics, Goethe University Frankfurt, 60438 Frankfurt am Main, Germany.}

\begin{abstract}
  Molecular self-organization driven by concerted many-body interactions produces the ordered structures that define both inanimate and living matter. Understanding the physical mechanisms that govern the formation of molecular complexes and crystals is key to controlling the assembly of nanomachines and new materials. We present an artificial intelligence (AI) agent that uses deep reinforcement learning and transition path theory to discover the mechanism of molecular self-organization phenomena from computer simulations. The agent adaptively learns how to sample complex molecular events and, on the fly, constructs quantitative mechanistic models.  By using the mechanistic understanding for AI-driven sampling, the agent closes the learning cycle and overcomes time-scale gaps of many orders of magnitude. Symbolic regression condenses the mechanism into a human-interpretable form. Applied to ion association in solution, gas-hydrate crystal formation, and membrane-protein assembly, the AI agent identifies the many-body solvent motions governing the assembly process, discovers the variables of classical nucleation theory, and reveals competing assembly pathways. The mechanistic descriptions produced by the agent are predictive and transferable to close thermodynamic states and similar systems.  Autonomous AI sampling has the power to discover assembly and reaction mechanisms from materials science to biology.
\end{abstract}

\maketitle

\section*{Introduction}

Understanding how generic yet subtly orchestrated interactions cooperate in the formation of complex structures is the key to steering molecular self-assembly \cite{pena-francesch2020BiosyntheticSelfhealingMaterials,vandriessche2018MolecularNucleationMechanisms}.  Molecular dynamics (MD) simulations promise us a detailed and unbiased view of self-organization processes.  Being based on accurate physical models, MD can reveal complex molecular reorganizations in a computer experiment with atomic resolution \cite{chung2015StructuralOriginSlow}.  However, the high computational cost of MD simulations can be prohibitive. Most collective self-organization processes are rare events that occur on time scales many orders of magnitude longer than the fast molecular motions limiting the MD integration step.  The system spends most of the time in metastable states and reorganizes during infrequent and rapid stochastic transitions between states. The transition paths (TPs) are the very special ``reactive'' trajectory segments that capture the reorganization process.  Learning molecular mechanisms from simulations requires computational power to be focused on sampling TPs \cite{dellago1998EfficientTransitionPathSampling} and distilling quantitative models out of them \cite{peters2006ObtainingReactionCoordinates}.  Due to the high dimensionality of configuration space, both sampling and information extraction are exceedingly challenging in practice. Here we address both challenges at once with an artificial intelligence (AI) agent that simultaneously builds quantitative mechanistic models of complex molecular events, validates the models on the fly, and uses them to accelerate the sampling by orders of magnitude compared to regular MD.

\section*{Results and Discussion}

\subsection*{Reinforcement learning of molecular mechanisms}

Statistical mechanics provides a general framework to obtain low-dimensional mechanistic models of self-organization events.  Here we focus on systems that reorganize between two states $A$ (assembled) and $B$ (disassembled), but generalizing to an arbitrary number of states is straightforward. Each TP connecting the two states contains a sequence of snapshots capturing the system during its reorganization. Consequently, the transition path ensemble (TPE) \emph{is} the mechanism at the highest resolution. Since the transition is effectively stochastic, quantifying its mechanism requires a probabilistic framework.  We define the committor $p_S(\mathbf{x})$ as the probability that a trajectory starting from a point $\mathbf{x}$ in configuration space enters state $S$ first, with $S=A$ or $B$, respectively, and $p_A(\mathbf{x}) + p_B(\mathbf{x}) = 1$ for ergodic dynamics.  As ideal reaction coordinates \cite{berezhkovskii2013DiffusionSplittingCommitment,e2006TheoryTransitionPaths}, the committors $p_A$ and $p_B=1-p_A$ report on the progress of the reaction $A\leftrightharpoons B$ and predict the trajectory fate in a Markovian way \cite{bolhuis2000ReactionCoordinatesBiomolecularIsomerization, best2005ReactionCoordinatesRates}.  The TPE and the committor jointly \emph{quantify} the mechanism.

Sampling the TPE and learning the committor function $p_B(\mathbf{x})$ are two outstanding and intrinsically connected challenges. Given that TPs are exceedingly rare in a very high-dimensional space, an uninformed search is futile. However, TPs converge near transition states \cite{best2005ReactionCoordinatesRates}, where the trajectory's evolution is yet uncommitted and $p_A(\mathbf{x})= p_B(\mathbf{x})= \sfrac{1}{2}$.  For Markovian dynamics the probability for a trajectory passing through $\mathbf{x}$ to be a TP satisfies $ P(\mathrm{TP}|\mathbf{x}) = 2 p_B (\mathbf{x}) (1 - p_B (\mathbf{x}))$, that is, the committor determines the probability of sampling a TP \cite{hummer2004TransitionPathsTransitionStates}. Therefore, learning the committor facilitates TP sampling and vice versa.  The challenges of information extraction and sampling are thus deeply intertwined. We thus tackle them simultaneously with the help of reinforcement learning \cite{mnih2015HumanlevelControlDeep,silver2017MasteringGameGo}.

\begin{figure*}[p] 
    \includegraphics[width=1\textwidth]{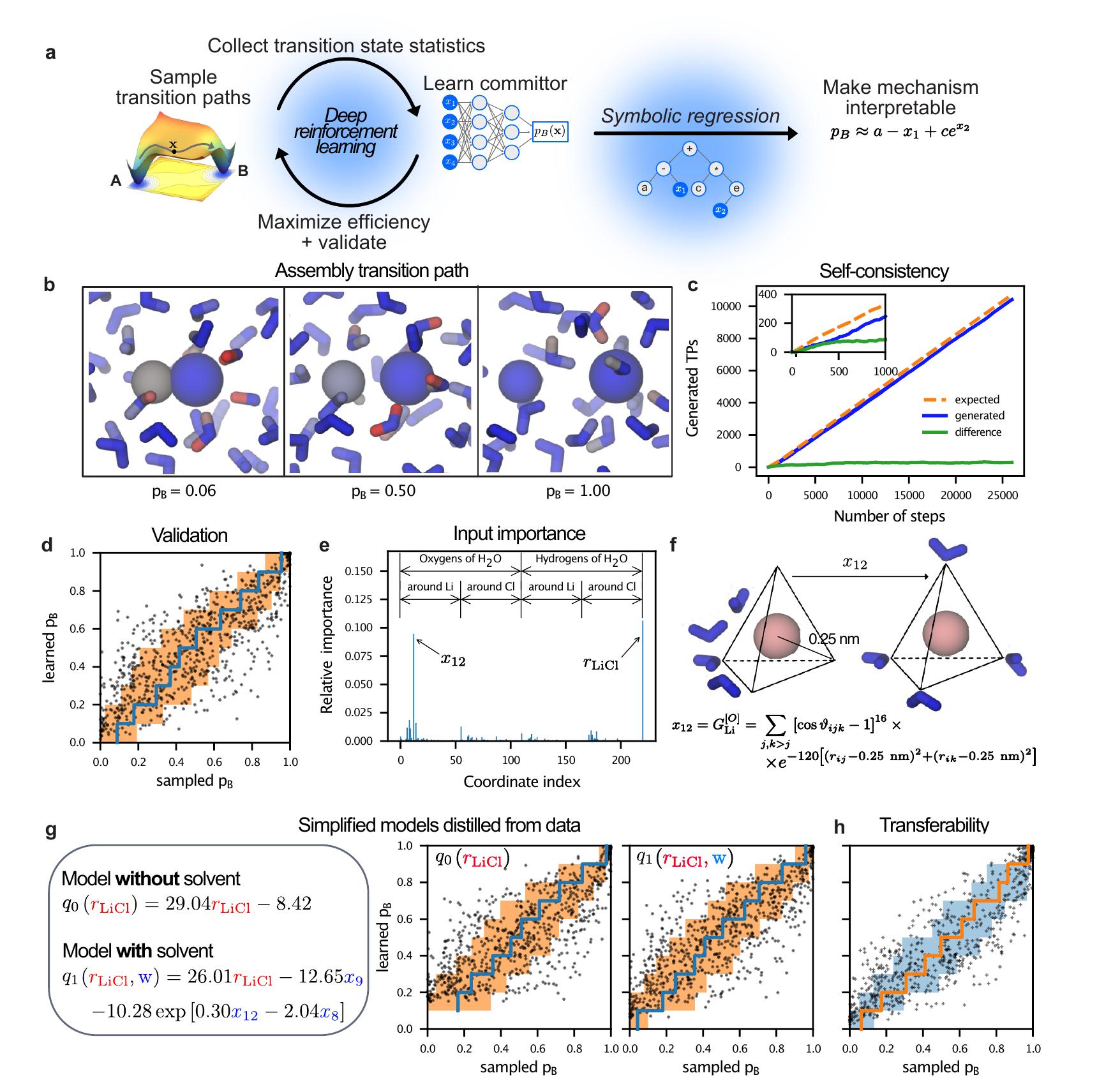}
    \renewcommand{\baselinestretch}{1.0}
    \caption{
    \textbf{Reinforcement learning of ion assembly mechanism.}
            \textbf{a}, Schematic of AI-driven reinforcement learning by path sampling.
            \textbf{b}, Snapshots along a TP showing the formation of a LiCl ion pair (right to left) in an atomistic MD simulation.
            Water is shown as sticks, Li$^+$ as a small sphere and Cl$^-$ as a large sphere.
            Atoms are colored according to their contribution to the reaction progress from low (blue) to high (red), as quantified by their contribution to the gradient of the reaction coordinate $q(\mathbf{x}|\mathbf{w})$. 
            \textbf{c}, Self-consistency. Cumulative counts of the generated (blue line) and expected (orange dashed line) number of transition events. The green line shows the cumulative difference between the observed and expected counts. The inset shows a zoom-in on the first 1000 iterations.
            \textbf{d}, Validation of the learned committor. Cross-correlation between the committor predicted by the trained network and the committor obtained by repeated sampling from molecular configurations on which the AI did not train. The average of the sampled committors (blue line) and their standard deviation (orange shaded) are calculated by binning along the predicted committor. 
            \textbf{e}, Input relevance for all 221 input coordinates used for deep learning. 
            \textbf{f}, Schematic depiction of the solvent reorientation around the Li$^+$ ion as reported by the most relevant symmetry function $x_{12}$.
            \textbf{g}, Simplified assembly models $q_0$ and $q_1$ obtained by symbolic regression at strict and gentle regularization, respectively.
            Scatter plots show independent validations of their accuracy.
            \textbf{h}, Transferability of the learned committor. Cross-correlation between sampled committors for 1M LiCl solution and predictions of committor trained on data for a single LiCl ion pair.}
    \label{fig:fig1}
\end{figure*}

\begin{figure}[htbp]
    \centering
    \includegraphics[width=0.9\columnwidth]{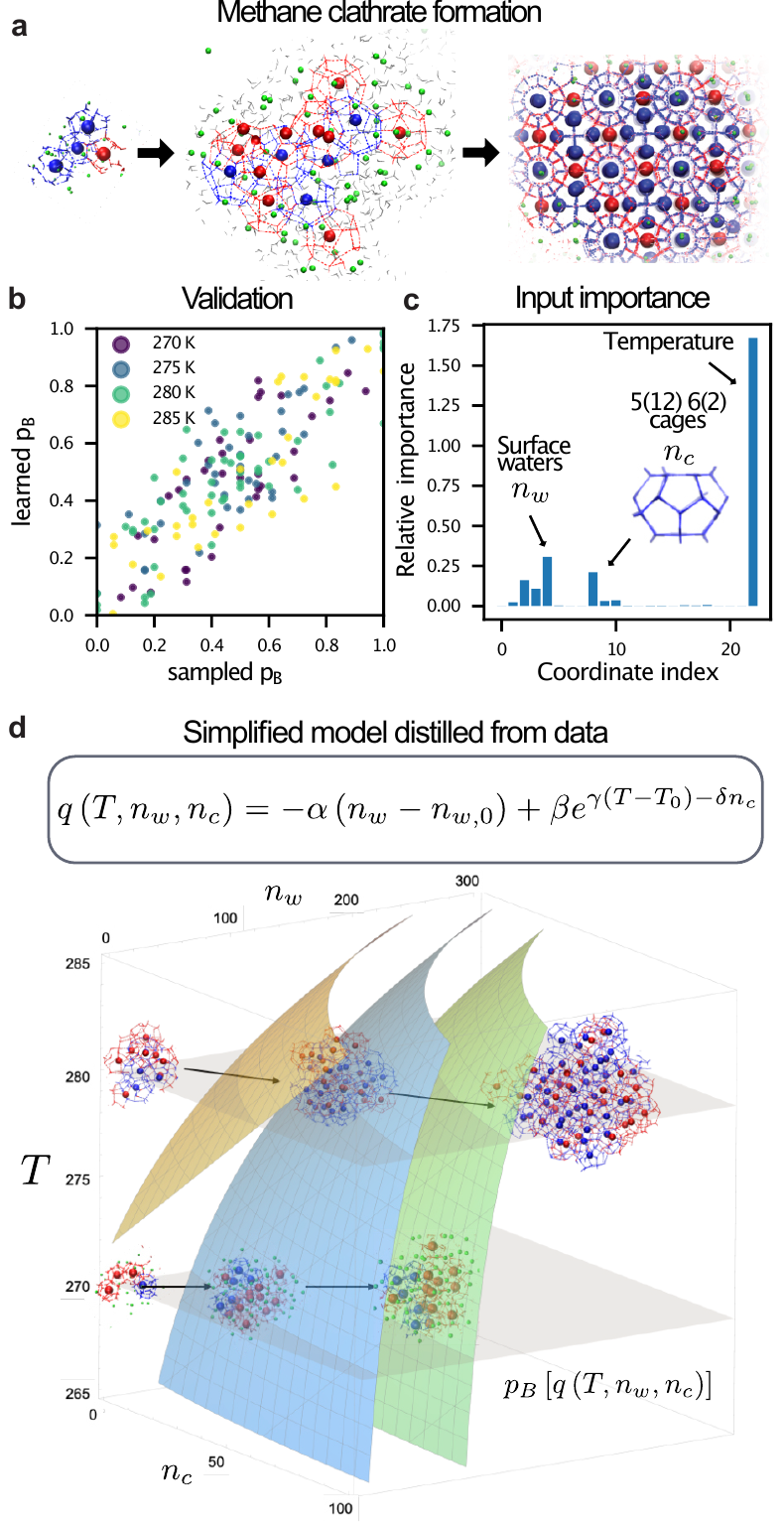}
    \renewcommand{\baselinestretch}{1.0}
    \caption{\textbf{Data-driven discovery of methane-clathrate nucleation mechanism.}
            \textbf{a}, Molecular configurations illustrating the nucleation process extracted from an atomistic MD simulation in explicit solvent. The nucleus forms in water, grows, and leads to the clathrate crystal. $5^{12}6^2$ (blue) and $5^{12}$ (red) water cages (lines) contain correspondingly colored methane molecules (spheres). Methane molecules near the growing solid nucleus are shown as green spheres, and water as gray sticks. Bulk water is not shown for clarity.  
            \textbf{b}, Validation of the learned committor. Cross-correlation between the committor predicted by the trained network and the committor obtained by repeated sampling from molecular configurations on which the AI agent did not train.
            \textbf{c}, Input importance analysis. The three most important input coordinates are annotated.
           \textbf{d}, Data-driven quantitative mechanistic model distilled by symbolic regression reveals a switch in nucleation mechanism. Analytical iso-committor surfaces for $n_{w,0}=2$, $T_0=270~\mathrm{K}$, $\alpha=0.0502$, $\beta=3.17$, $\gamma=0.109~\mathrm{K^{-1}}$, $\delta=0.0149$ (left to right: $p_B=1/( 1 + e^{-4} )$, $1/2$, $1/( 1 + e^4 )$).
            The structural insets illustrate the two competing mechanisms at low and high temperature.} 
    \label{fig:fig2}
\end{figure}

  We designed an AI agent that learns how to sample the TPE of rare events in complex many-body systems and at the same time learns their committor function by repeatedly running virtual experiments (Fig.~\ref{fig:fig1}a). In each experiment, the AI agent selects a point $\mathbf{x}$ from which to shoot trajectories---propagated according to the dynamics of the physical model---to generate TPs. After repeated shots from different points $\mathbf{x}$, the agent compares the number of generated TPs with the expected number based on its knowledge of the committor.  Only if the prediction is poor, the AI agent retrains the model on the outcome of all virtual experiments, which prevents over-fitting.  As the agent becomes better at predicting the outcome of the virtual experiment, it becomes more efficient at sampling TPs by choosing initial points near transition states, i.e., according to $P(\mathrm{TP}|\mathbf{x})$.

The AI agent learns from its repeated attempts by using deep learning in a self-consistent way. Here, we model the committor $p_{B}(\mathbf{x})=1/(1+e^{-q(\mathbf{x|w})})$ with a neural network\cite{ma2005AutomaticMethodIdentifying} $q(\mathbf{x|w})$ of weights $\parameters$. Note that $\mathbf{x}$ interchangeably denotes a vector of features and the configuration represented by this vector. In each attempt to generate a TP, the agent propagates two trajectories, one forward and one backward in time, by running MD simulations \cite{dellago1998EfficientTransitionPathSampling}. In case of success, one trajectory first enters state $A$ and the other $B$, forming a new TP. However, the agent learns from both successes and failures of this Bernoulli coin-toss process.  The negative log-likelihood \cite{peters2006ObtainingReactionCoordinates} of $k$ attempts leads to the loss function $l(\parameters|\boldsymbol{\theta}) = \sum_{i=1}^{k} \log ( 1+e^{s_i q(\mathbf{x}_i|\mathbf{w})} )$, where $s_i=1$ if trajectory $i$ initiated from $\mathbf{x}_i$ enters $A$ first, and $s_i=-1$ if it enters $B$ first. The training set $\boldsymbol{\theta}$ contains shooting points $\mathbf{x}_i$ and outcomes $s_i$.  By training the network $q(\mathbf{x}|\mathbf{w})$ to minimize the loss $l(\mathbf{w}|\boldsymbol{\theta})$, the agent obtains a maximum likelihood estimate of the committor \cite{peters2006ObtainingReactionCoordinates}.

We use machine learning also to condense the learned molecular mechanism into a human-interpretable form (Fig.~\ref{fig:fig1}a). The trained network is efficient to evaluate, differentiable, and enables sophisticated analysis methods \cite{vanden-eijnden2008AssumptionsUnderlyingMilestoning}. To provide physical insight, symbolic regression \cite{schmidt2009DistillingFreeFormNatural} generates simple models that quantitatively reproduce the committor encoded in the network. First, a sensitivity analysis of the trained network identifies a small subset $\mathbf{z}\subset\mathbf{x}$ of all input coordinates $\mathbf{x}$ that controls the quality of the prediction by the network.  Then, symbolic regression distills mathematical expressions $q_{\mathrm{sr}}(\mathbf{z}) \approx q(\mathbf{x|w})$ by using a genetic algorithm that searches both functional and parameter spaces with loss $l(\mathbf{w}|\boldsymbol{\theta})$ and training set $\boldsymbol{\theta}$.

\subsection*{AI discovers many-body solvent coordinate in ion assembly}
The formation of ion pairs in water is a paradigmatic assembly process controlled by many-body interactions in the molecular environment---the solvent, in this case---and a  model of chemical reactions in condensed phase. Even though MD can efficiently simulate the process, the collective reorganization of water molecules challenges the formulation of quantitative mechanistic models to this day \cite{ballard2012MechanismIonicDissociation} (Fig.~\ref{fig:fig1}b).

Our AI agent quickly learned how to sample the formation of lithium (Li$^+$) and chloride (Cl$^-$) ion pairs in water (Fig.~\ref{fig:fig1}b, c). As input, the network uses the interionic distance $r_{\mathrm{LiCl}}$ and 220 molecular features $x_1,\dots,x_{220}$ that describe the angular arrangement of water oxygen and hydrogen atoms at a specific distance from each ion \cite{behler2007GeneralizedNeuralNetworkRepresentation} (Fig.~\ref{fig:fig1}e, f). These coordinates provide a general representation of the system that is invariant with respect to physical symmetries and exchange of atom labels. After the first 500 iterations, the predicted and observed numbers of TPs agree (Fig.~\ref{fig:fig1}c).  We further validated the learned committor function by checking its predictions against independent simulations.  From 763 configurations not used in AI training, we initiated 500 independent simulations each and estimated the sampled committor $p_B$ as the fraction of trajectories first entering the unbound state. Predicted and sampled committors are in quantitative agreement (Fig.~\ref{fig:fig1}d).

The input importance analysis of the trained network reveals the critical role played by solvent rearrangement. As the most important of the 220 molecular features used to describe the solvent, $x_{12}$ quantifies oxygen anisotropy at a radial distance of 0.25 nm from Li$^+$ (Fig.~\ref{fig:fig1}f).  For successful ion-pair assembly, these inner-shell water molecules need to open up space for the incoming Cl$^-$.  The importance of inner-shell water rearrangement is consistent with a visual analysis that highlights atoms in a TP according to their contribution to the committor gradient (Fig.~\ref{fig:fig1}b).

Symbolic regression provides quantitative and interpretable models of the ion-pair assembly mechanism. A large complexity penalty produces a simple model $q_0(r_{\mathrm{LiCl}})$ as a function of the inter-ionic distance only---the standard choice to study this process (Fig.~\ref{fig:fig1}g).  In a validation test, we found this model to be predictive only for large inter-ionic distances. Close to the bound state, where the detailed geometry of the water solvent controls the process, the distance-only model fails. A more complex model, $q_1(r_{\mathrm{LiCl}}, x_8, x_9, x_{12})$, that integrates the most critical solvent coordinates is accurate for all distances (Fig.~\ref{fig:fig1}g).

Counter to a common concern for AI models, the learned committor is transferable to a similar but not identical system. We validated the learned committor for 724 configurations drawn from an additional simulation of a 1M concentrated LiCl solution. Even though AI trained on a system containing a single ion pair, it correctly predicted committors for a system on which it never trained (Fig.~\ref{fig:fig1}h).

\subsection*{AI discovers variables of classical nucleation theory for gas-hydrate crystal formation}

At low temperature and high pressure, a liquid mixture of water and methane organizes into a gas hydrate, an ice-like solid \cite{walsh2009SimulationsSpontaneousMethaneHydrateNucleation}.
In this phase transition, water molecules assemble into an intricate crystal lattice with regularly spaced cages filled by methane (Fig.~\ref{fig:fig2}a).
Despite commercial relevance in natural gas processing,
the mechanism of gas-hydrate formation remains poorly understood, complicated by the many-body character of the nucleation process and the competition between different crystal forms \cite{walsh2009SimulationsSpontaneousMethaneHydrateNucleation}.
Studying the nucleation mechanism is challenging for experiments and, due to the exceeding rarity of the events, impossible in equilibrium MD.

\begin{figure*}[tbh]
    \centering
    \includegraphics[width=1\textwidth]{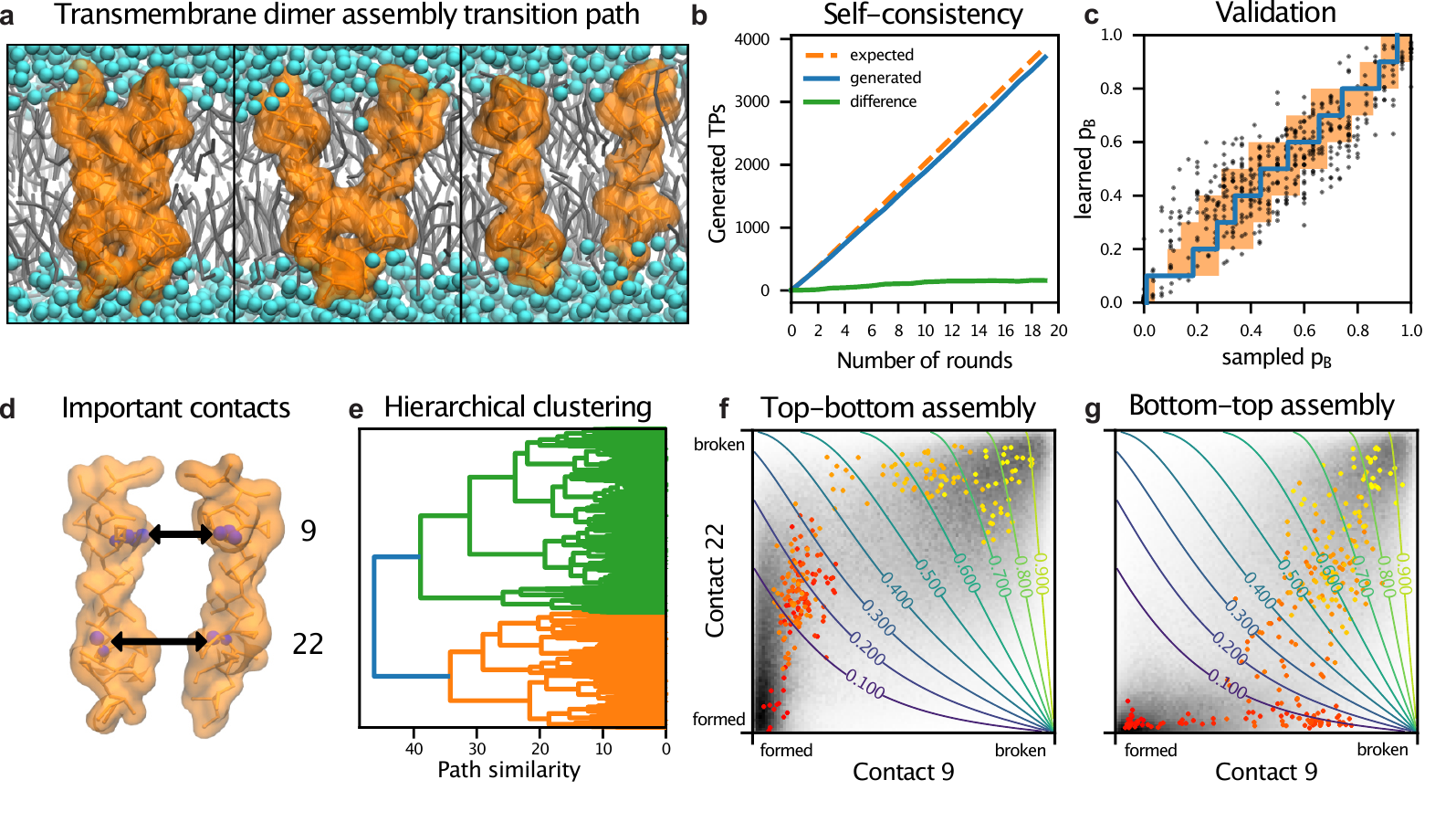}
    \caption{\textbf{Competing pathways of transmembrane dimer assembly in lipid membrane.}
    \textbf{a}, Snapshots during a Mga2 dimerization event (right to left).
    The transmembrane helices are shown as orange surfaces, the lipid molecules in grey, and water in cyan.
    \textbf{b}, Self-consistency. Cumulative counts of the generated (blue line) and expected (orange dashed line) number of transitions. The green curve shows the cumulative difference between the observed and expected counts.
    \textbf{c}, Validation of the learned committor. Cross-correlation between the committor predicted by the trained network and the committor obtained by repeated sampling from molecular configurations on which the AI did not train. The average of the sampled committors (blue line) and their standard deviation (orange shaded) are calculated by binning along the predicted committor. 
    \textbf{d}, Schematic representation of the two most relevant coordinates, the interhelical contacts at position 9 and 22.
    \textbf{e}, Hierarchical clustering of all TPs. Dendrogram as a function of TP similarity (dynamic time warping, see Methods) calculated in the plane defined by contacts 9 and 22.
    \textbf{f}, \textbf{g}, Path density (gray shading) for the two main clusters (f, green; and g, orange in panel E) calculated in the plane defined by contacts 9 and 22. For each cluster, one representative TP is shown from unbound (yellow) to bound (red). The isolines of the committor, as predicted by the symbolic regression $q_B(x_9, x_{22})$, are shown as labelled solid lines.}
    \label{fig:fig3}
  \end{figure*}

  Within hours of computing time, the AI extracted the nucleation mechanism from 2225 TPs showing the formation of methane clathrates, corresponding to a total of 445.3~$\mu$s of simulated dynamics. The trajectories were recently produced by extensive transition path sampling (TPS) simulations at four different temperatures, and provided a pre-existing training set for our AI \cite{arjun2019UnbiasedAtomisticInsight}. We described molecular configurations by using 22 features commonly used to describe nucleation processes (SI Table \ref{tab:clathrate_descriptors}). We considered the temperature $T$ at which a TP was generated as an additional feature, and trained the AI on the cumulative trajectories. We showed that the learned committor as a function of temperature is accurate by validating its predictions for 160 independent configurations (Fig.~\ref{fig:fig2}b). By leaving out data at $T= 280$~K or $285$~K in the training, we show that the learned committor satisfactorily interpolates and extrapolates to thermodynamic states not sampled (SI Fig. \ref{fig:SI_clathrates_leave_one_T_out}).

Temperature $T$ is the most critical factor for the outcome of a simulation trajectory, followed by the number $n_w$ of surface water molecules and the number $n_c$ of $5^{12}6^2$ cages with 12 pentagons and two hexagons (Fig.~\ref{fig:fig2}c). All three play an essential role in the classical theory of homogeneous nucleation \cite{arjun2019UnbiasedAtomisticInsight}. The activation free energy $\Delta G$ for nucleation is determined by the size of the growing nucleus, parametrized by the amount of surface water and---in case of a crystalline structure---the number of $5^{12}6^2$ cages. Temperature determines, through the degree of supersaturation, the size of the critical nucleus, the nucleation free energy barrier height and the rate.

Symbolic regression distilled a mathematical expression revealing a temperature dependent switch in the nucleation mechanism. The mechanism is quantified by $q( n_w, n_c, T )$ (Fig.~\ref{fig:fig2}d; SI Table \ref{tab:clathrate_descriptors}). At low temperatures, the size of the nucleus alone decides on growth. At higher temperatures, the number of $5^{12}6^2$ water cages gains in importance, as indicated by curved iso-committor surfaces (Fig.~\ref{fig:fig2}d). This mechanistic model, generated in an autonomous and completely data-driven way, reveals the switch from amorphous growth at low temperatures to crystalline growth at higher temperatures
\cite{arjun2019UnbiasedAtomisticInsight,jacobson2010AmorphousPrecursorsNucleation}.

\subsection*{AI reveals competing pathways for membrane-protein complex assembly}

Membrane protein complexes play a fundamental role in the organization of living cells. 
Here we investigate the assembly of the transmembrane homodimer of the saturation sensor Mga2 in a lipid bilayer in the quasi-atomistic Martini representation (Fig.~\ref{fig:fig3}a) \cite{covino2016EukaryoticSensorMembraneSaturation}. In extensive equilibrium MD simulations, the spontaneous association of two Mga2 transmembrane helices has been observed, yet no dissociation occurred in approximately 3.6 milliseconds (equivalent to more than six months of calculations) \cite{covino2016EukaryoticSensorMembraneSaturation}. 

Our reinforcement learning AI approach is naturally parallelizable, which enabled us to sample nearly 4,000 dissociation events in 20 days on a parallel supercomputer (Fig.~\ref{fig:fig3}b). MD time integration incurs the highest computational cost. However, a single AI agent can simultaneously perform virtual experiments on an arbitrary number of copies of the physical model (by guiding parallel Markov chain Monte Carlo sampling processes), and learn from all of them by training on the cumulative outcomes.

We featurized molecular configurations using contacts between corresponding residues along the two helices and included, for reference, a number of hand-tailored features describing the organization of lipids around the proteins \cite{chiavazzo2017IntrinsicMapDynamics} (SI Table \ref{tab:MGA2_descriptors}). We validated the model against committor data for 548 molecular configurations not used in training, and found the predictions to be accurate across the entire transition region between bound and unbound states (Fig.~\ref{fig:fig3}c).

In a remarkable reduction of dimensionality, symbolic regression achieved an accurate representation of the learned committor as a simple function of just two amino-acid contacts (Fig.~\ref{fig:fig3}d; SI Table \ref{tab:MGA2_2d_symregs}). We projected all sampled TPs on the plane defined by these two contacts, calculated the distances between them, and performed a hierarchical trajectory clustering (Fig.~\ref{fig:fig3}e). TPs organize in two main clusters that differ in the order of events during the assembly process---starting from the top (Fig.~\ref{fig:fig3}f) or the bottom (Fig.~\ref{fig:fig3}g)---revealing two competing assembly pathways. Unexpectedly \cite{chiavazzo2017IntrinsicMapDynamics}, helix-dimer geometry alone predicts assembly progress, which implies that the lipid ``solvent'' is implicitly encoded, unlike the water solvent in ion-pair formation \cite{ballard2012MechanismIonicDissociation}.

\section*{Methods}
\subsection*{Maximum likelihood estimation of the committor function}
The committor $p_B(\mathbf{x})$ is the probability that a trajectory initiated at configuration $\mathbf{x}$ with Maxwell-Boltzmann velocities reaches the (meta)stable state $B$ before reaching $A$.
Trajectory shooting thus constitutes a Bernoulli process.
We expect to observe $n_A$ and $n_B$ trajectories to end in $A$ and $B$, respectively, with binomial probability $p( n_A, n_B|\mathbf{x}) = \binom{n_A + n_B}{n_A} (1 - p_B(\mathbf{x}))^{n_A} p_B(\mathbf{x})^{n_B}$.
For $k$ shooting points $\mathbf{x}_{i}$, the combined probability defines the likelihood $\mathcal{L}=\prod_{i=1}^{k} p( n_A(i), n_B(i)|\mathbf{x}_{i})$.
Here we ignore the correlations that arise in fast inertia-dominated transitions for trajectories shot off with opposite initial velocities \cite{hummer2004TransitionPathsTransitionStates,ballard2012MechanismIonicDissociation}. We model the unknown committor with a parametric function and estimate its parameters $\parameters$ by maximizing the likelihood $\mathcal{L}$ \cite{ma2005AutomaticMethodIdentifying,peters2006ObtainingReactionCoordinates}.
We ensure that $0\le p_B(\mathbf{x})\le 1$ by writing the committor in terms of a sigmoidal activation function, $p_B[q(\mathbf{x}| \parameters)]=1/(1+\exp[-q(\mathbf{x}| \parameters)])$. Here we model the log-probability $q(\mathbf{x}| \parameters)$ using a neural network \cite{ma2005AutomaticMethodIdentifying} and represent the configuration with a vector $\mathbf{x}$ of features.
For $N>2$ states $S$, the multinomial distribution provides a model for $p(n_1, n_2, ..., n_N|\mathbf{x})$,
and writing the committors to states $S$ in terms of the softmax activation function ensures normalization, $\sum_{S=1}^{N} p_S = 1$.
The loss function $l(\mathbf{w}|\boldsymbol{\theta})$ used in the training
is the negative logarithm of the likelihood $\mathcal{L}$.

\subsection*{Training points from transition path sampling}
TPS
\cite{dellago1998EfficientTransitionPathSampling,bolhuis2002TRANSITIONPATHSAMPLING}
is a powerful Markov chain Monte Carlo method in (transition) path space to sample the TPE. The two-way shooting move is an efficient proposal move in TPS \cite{dellago1998EfficientTransitionPathSampling}. It consists of randomly selecting a shooting point $\mathbf{x}_{\mathrm{SP}}$ on the current TP $\chi$ according to probability $p_{\mathrm{sel}}(\mathbf{x}_{\mathrm{SP}}|\chi)$, drawing random Maxwell-Boltzmann velocities, and propagating two trial trajectories from $\mathbf{x}_{\mathrm{SP}}$ until they reach either one of the states. Because one of the trial trajectories is propagated after first inverting all momenta at the starting point, i.e., it is propagated backwards in time, a continuous TP can be constructed if both trials end in different states. Given a TP $\chi$, a new TP $\chi'$ generated by two-way shooting is accepted into the Markov chain with probability \cite{jung2017TransitionPathSampling}
$p_{\mathrm{acc}}(\chi'| \chi) = \min( 1, {p_{\mathrm{sel}}(\mathbf{x}_{\mathrm{SP}}|\chi')}/{p_{\mathrm{sel}}(\mathbf{x}_{\mathrm{SP}}|\chi)} )$.
If the new path is rejected, $\chi$ is repeated.

Knowing the committor it is possible to increase the rate at which TPs are generated by biasing the selection of shooting points towards the transition state ensemble \cite{jung2017TransitionPathSampling}, i.e., regions with high reactive probability $p(\mathrm{TP}|\mathbf{x})$. For the two-state case, this is equivalent to biasing towards the $p_B(\mathbf{x})=\sfrac{1}{2}$ isosurface defining the transition states with $q(\mathbf{x})=0$. To construct an algorithm which selects new shooting points biased towards the current best guess for the transition state ensemble and which iteratively learns to improve its guess based on every newly observed shooting outcome, we need to balance exploration with exploitation. To this end, we select the new shooting point $\mathbf{x}$ from the current TP $\chi$ using a Lorentzian distribution centered around the transition state ensemble,
$p_{\mathrm{sel}}(\mathbf{x}|\chi) = 1/{\sum\limits_{\mathbf{x}^{\prime} \in \chi}[(q(\mathbf{x})^2 + \gamma^2)/(q(\mathbf{x}^\prime)^2 + \gamma^2)]},$
where larger values of $\gamma$ lead to an increase of exploration. The Lorentzian distribution provides a trade-off between production efficiency and the occasional exploration away from the transition state, which is necessary to sample alternative reaction channels.

\subsection*{Real-time validation of committor model prediction}
The relation between the committor and the transition probability \cite{hummer2004TransitionPathsTransitionStates} enables us to calculate the expected number of TPs generated by shooting from a configuration $\mathbf{x}$. We validate the learned committor on-the-fly by estimating the expected number of transitions before shooting from a configuration and comparing it with the observed shooting result.
The expected number of transitions $n_{\mathrm{TP}}^{\mathrm{exp}}$ calculated over a window containing the $k$ most recent two-way shooting \cite{dellago1998EfficientTransitionPathSampling} attempts is
$n_{\mathrm{TP}}^{\mathrm{exp}} = \sum_{i=1}^{k} 2(1- p_B(\mathbf{x}_{i},i)) p_B(\mathbf{x}_{i},i)$,
where $p_B(\mathbf{x}_{i},i)$ is the committor estimate for trial shooting point $\mathbf{x}_{i}$ at step $i$ before observing the shooting result.
We initiate learning when the predicted ($n_{\mathrm{TP}}^{\mathrm{exp}}$) and actually generated number of TPs ($n_\mathrm{TP}^{\mathrm{gen}}$) differ. We define an efficiency factor,
$\alpha_{\mathrm{eff}} = \min{(1, (1 - n_{\mathrm{TP}}^{\mathrm{gen}} / n_{\mathrm{TP}}^{\mathrm{exp}})^2)}$,
where a value of zero indicates perfect prediction.
By training only when necessary we avoid overfitting. Here we use $\alpha_{\mathrm{eff}}$ to scale the learning rate in the gradient descent algorithm. Additionally, no training takes place if $\alpha_{\mathrm{eff}}$ is below a certain threshold (specified further below for each system).

\subsection*{Distilling explicit mathematical expressions for the committor}
In any specific molecular event, we expect that only few of the many degrees of freedom actually control the transition. We identify the inputs to the committor model that have the largest role in determining its output after training. To this end we first calculate a reference loss, $l_{\mathrm{ref}}= l(\parameters, \boldsymbol{\theta})$, over the unperturbed training set to compare to the values obtained by perturbing every input one by one \cite{kemp2007DeterminingRelativeInputImportance}. We then average the loss  $l(\parameters, \Tilde{\boldsymbol{\theta}}_{i})$ over $\ge 100$ perturbed training sets $\Tilde{\boldsymbol{\theta}}_{i}$ with randomly permuted  values of the input coordinate $i$ in the batch dimension. The average loss difference $\Delta l_{i} = \left\langle l(\parameters, \Tilde{\boldsymbol{\theta}}_{i}) \right\rangle - l_{\mathrm{ref}}$ is large if the $i$th input strongly influences the output of the trained model, i.e., it is relevant for predicting the committor.

In the low-dimensional subspace consisting of only the most relevant inputs $\mathbf{z}$, symbolic regression generates compact mathematical expressions that approximate the full committor. Our implementation of symbolic regression is based on the python package dcgpy \cite{izzo2020DcgpDifferentiableCartesian} and uses a genetic algorithm with a ($N$ + 1) evolution strategy. In every generation, $N$ new expressions are generated through random changes to the mathematical structure of the fittest expression of the parent generation. A gradient based optimization is subsequently used to find the best parameters for every expression. The fittest expression is then chosen as parent for the next generation. The fitness of each trial expression $p_{B}(\mathbf{z})$
is measured by $l_{\mathrm{sr}}(p_{B}|\boldsymbol{\theta})\equiv-\log \mathcal{L}[p_{B}(\mathbf{z}_{\mathrm{sp}})]+\lambda C$,
where we added the regularization term $\lambda C$ to the log-likelihood
in order to keep
expressions simple and avoid over-fitting. Here $\lambda>0$ and $C$
is a measure of the complexity of the trial expression, estimated in our case by the number of mathematical operations.

\subsection*{Assembly of LiCl in water}
We investigated the formation of lithium chloride ion pairs in water to asses the ability of our AI agent to accurately learn the committor for transitions that are strongly influenced by solvent degrees of freedom. We used two different system setups, one consisting of only one ion pair in water and one with a number of ions corresponding to a 1 molar (1M) concentration.

All MD simulations were carried out in cubic simulation boxes using the Joung and Cheatham forcefield \cite{joung2008DeterminationAlkaliHalide} together with TIP3P \cite{jorgensen1983ComparisonSimplePotentialWater} water. The 1M simulation box contained 37 lithium and 37 chloride ions, solvated with 2104 TIP3P water molecules, while the other box contained the single ion pair solvated with 370 TIP3P water molecules. We used the openMM MD engine \cite{eastman2017OpenMM7} to propagate the equations of motion in time steps of $\Delta t=2~\mathrm{fs}$ with a velocity Verlet integrator with velocity randomization \cite{sivak2014TimeStepRescaling}
from the python package openmmtools. After an initial NPT equilibration at $P=1~\mathrm{bar}$ and $T=300~\mathrm{K}$, all production simulations were performed in the NVT-Ensemble at a temperature of $T=300~\mathrm{K}$. The friction was set to $1/\mathrm{ps}$.
Non-bonded interactions were calculated using a particle mesh Ewald
scheme \cite{essmann1995SmoothParticleMeshEwald} with a cutoff of 1 nm and an error tolerance
of $0.0005$.
In TPS, the fully assembled and disassembled states were defined according to interionic distances $r_{\mathrm{LiCl}} \le 0.23 ~\mathrm{nm}$ and  $r_{\mathrm{LiCl}} \ge 0.48 ~\mathrm{nm}$, respectively.

The committor of a configuration is invariant under global translations and rotations in the absence of external fields, and it is additionally invariant with respect to permutations of identical particles. 
We therefore chose to transform the systems coordinates from the Cartesian space to a representation that incorporates the physical symmetries of the committor. To achieve an almost lossless transformation, we use the interionic distance to describe the solute and we adapted symmetry functions to describe the solvent configuration \cite{behler2011AtomcenteredSymmetryFunctions}. Symmetry functions have been developed originally to describe molecular structures in neural network potentials \cite{behler2007GeneralizedNeuralNetworkRepresentation,behler2014RepresentingPotentialEnergy}, but have also been successfully used to detect and describe different phases of ice in atomistic simulations \cite{geiger2013NeuralNetworksStructureDetection}. These functions describe the  environment surrounding a central atom by summing over all identical particles at a given radial distance. The $G_{i}^2$ type of symmetry function quantifies the density of solvent molecules around a solute atom $i$ in a shell centered at $r_s$,
\begin{equation*}
G_{i}^{2}=\sum_{j}e^{-\eta(r_{ij}-r_{s})^{2}}f_{c}(r_{ij}),\label{eq: G2}
\end{equation*}
where the sum runs over all solvent atoms $j$ of a specific atom type, $r_{ij}$ is the distance between the central atom $i$ and atom $j$ and $\eta$ controls the width of the shell.
The function $f_{c}(r)$ is a Fermi cutoff defined as:
\begin{equation*}
f_{c}(r)=\begin{cases}
\left[1+\exp(\alpha_{c}(r-r_{\mathrm{cut}}-1/\sqrt{\alpha_{c}}))\right]^{-1} & r\leq r_{\mathrm{cut}}\\
0 & r>r_{\mathrm{cut}}
\end{cases},
\end{equation*}
which ensures that the contribution of distant solvent atoms vanishes. The scalar parameter $\alpha_c$ controls the steepness of the cutoff.
The $G_{i}^5$ type of symmetry function additionally probes the angular distribution
of the solvent around the central atom $i$,
\begin{alignat}{1}\label{eq:G5}
G_{i}^{5}=\sum_{j,k>j}\left(1+\lambda\cos\vartheta_{ijk}\right)^{\zeta} & e^{-\eta\left[(r_{ij}-r_{s})^{2}+(r_{ik}-r_{s})^{2}\right]}\nonumber\\\times
f_{c}(r_{ik})f_{c}(r_{ij}),\nonumber 
\end{alignat}
where the sum runs over all distinct solvent atom pairs, $\vartheta_{ijk}$ is
the angle spanned between the two solvent atoms and the central solute
atom, the parameter $\zeta$ is an even number that controls the
sharpness of the angular distribution, and  $\lambda=\pm 1$ sets the location of
the minimum with respect to $\vartheta_{ijk}$ at $\pi$ and $0$, respectively. See SI Table \ref{tab:LiCl_sf_params} for the parameter combinations used. We scaled all inputs to lie approximately in the range $[0,1]$ to increase the numerical stability of the training. In particular, we normalized the symmetry functions by dividing them by the expected average number of atoms (or atom pairs) for an isotropic distribution in the probing volume (see SI for mathematical expressions of the normalization constants as a function of the parameters).

Due to the expectation that most degrees of freedom of the system do not control the transition, we designed neural networks that progressively filter out irrelevant inputs and build a highly non-linear function of the remaining ones. We therefore used a pyramidal stack of five residual units \cite{he2015DeepResidualLearning,he2016IdentityMappingsDeep}, each with four hidden layers. The number of hidden units per layer is reduced by a constant factor $f=(10/221)^{1/4}$
after every residual unit block and decreases from 221 in the first unit to 10 in the last. Additionally, a dropout of $0.1 f^i$, where $i$ is the residual unit index ranging from $0$ to $4$, is applied after every residual block. Optimization of the network weights is performed using the Adam gradient descent \cite{kingma2017AdamMethodStochastic}. Training was performed after every third TPS Monte Carlo step for one epoch with a learning rate of $lr=\alpha_{\mathrm{eff}}10^{-3}$, if $lr \ge 10^{-4}$.
The expected efficiency factor $\alpha_{\mathrm{eff}}$ was calculated over a window of $k=100$ TPS steps. We performed all deep learning with custom written code based on keras \cite{chollet2015Keras}. The TPS simulations were carried out using a customized version of openpathsampling \cite{swenson2019OpenPathSampling_1Basics,swenson2019OpenPathSampling_2Customizing} together with our own python module.
We selected the five most relevant coordinates for symbolic regression runs. We regularized the produced expressions by penalizing the total number of elementary mathematical operations with $\lambda=10^{-6}$
and $\lambda=10^{-7}$.
The contributions of each atom in the system to the committor (Fig. \ref{fig:fig1}b) was calculated as the magnitude of the gradient of the reaction coordinate $q(\mathbf{x})$ with respect to its Cartesian coordinates. 
All gradient magnitudes were scaled with the inverse atom mass.

\subsection*{Nucleation of methane clathrates}
We modelled water with the TIP4P/Ice model \cite{tip4pice} and methane with the united atom Lennard-Jones interactions ($\epsilon$ = 1.22927 kJ/mol and $\sigma$ = 3.700 \AA), which reproduce experimental measurements well \cite{conde2010determining}. MD simulations were performed using OpenMM 7.1.1~\cite{eastman2017OpenMM7}, integrating the equations of motion with the Velocity Verlet with velocity randomisation (VVVR) integrator from openmmtools~\cite{sivak2014TimeStepRescaling}. The integration time step was set to 2 fs. Hydrogen bond lengths were constrained \cite{hess1997LINCSLinearConstraint}. The van der Waals cutoff distance was 1 nm.  Long range interactions were handled by the Particle Mesh Ewald technique. The MD simulations were performed in the NPT ensemble using the VVVR thermostat (frequency of 1 ps) and a Monte Carlo barostat (frequency of 4 ps). TPS simulations were performed with the OpenPathSampling package \cite{swenson2019OpenPathSampling_1Basics,swenson2019OpenPathSampling_2Customizing} using the CUDA platform of OpenMM on NVIDIA GeForce GTX TITAN 1080Ti GPUs. The saving frequency of the frames was every 100 ps. TPS and committor simulations were carried out at four different temperatures $T=270$ K, 275 K, 280 K and 285 K (see SI Table \ref{tab:clathrates_dataset} for details). The committor values, which were used only for the validation, were obtained by shooting between 6 and 18 trajectories per configuration.
The disassembled (liquid state) and assembled (solid) states were defined in terms of the mutually coordinated guest (MCG) numbers as in Ref.
\cite{arjun2019UnbiasedAtomisticInsight}.

We used 24 different features to describe size, crystallinity, structure, and composition of the growing methane-hydrate crystal  nucleus (SI Table  \ref{tab:clathrate_descriptors}). In addition to the features describing molecular configurations we used temperature as an input to the neural networks and the symbolic regression. In a pyramidal feed forward network with 9 layers, we reduced the number of units per layer from 25 at the input to two
in the last hidden layer. The network was trained on the existing TPS data for all temperatures, leaving out 10 \% of the shooting points as test data. We stopped the training after the loss on the test set did not decrease for 10000 epochs and used the model with the best test loss. We used the three most relevant coordinates as inputs for symbolic regression runs with a penalty on the total number of elementary mathematical operations using $\lambda=10^{-5}$.

\subsection*{Mga2 transmembrane dimer assembly in lipid membrane}
We used the coarse-grained Martini force field (v2.2) \cite{marrink2004CoarseGrainedModel,marrink2007MARTINIForceField,monticelli2008MARTINICoarseGrainedProteins,dejong2013ImprovedParametersMartini} to describe the assembly of the alpha-helical transmembrane homodimer Mga2. All MD simulations were carried out with gromacs v4.6.7 \cite{berendsen1995GROMACSMessagepassingParallel,hess2008GROMACS4,pronk2013GROMACS4.5,abraham2015GROMACSHighPerformance} with an integration timestep of $\Delta t=0.02~\mathrm{ps}$, using a cubic simulation box containing the two identical 30 amino acid long alpha helices in a lipid membrane made of 313 POPC molecules. The membrane spans the box in the $xy$ plane and was solvated with water (5348 water beads) and NaCl ions corresponding to a concentration of 150~mM (58 Na$^+$, 60 Cl$^-$). A reference temperature of $T=300~\mathrm{K}$ was enforced using the v-rescale thermostat \cite{bussi2007CanonicalSamplingVelocity} with a coupling constant of $1~\mathrm{ps}$ separately on the protein, the membrane, and the solvent. A pressure of $1~\mathrm{bar}$ was enforced separately in the $xy$ plane and in $z$ using a semiisotropic Parrinello-Rahman barostat \cite{parrinello1981PolymorphicTransitionsSingle} with a time constant $12~\mathrm{ps}$ and compressibility $3\cdot10^{-4}~\mathrm{bar}^{-1}$.

To describe the assembly of the Mga2 homodimer we used 28 interhelical pairwise distances between the backbone beads of the two helices together with the total number of interhelical contacts, the distance between the helix centers of mass, and a number of hand-tailored features describing the organization of lipids around the two helices (SI Table \ref{tab:MGA2_descriptors}). To ensure that all network inputs lie approximately in $[0,1]$, we used the sigmoidal function $f(r) = (1 - {r}/{R_0})^6 / (1 - {r}/{R_0})^{12}$ with $R_0 = 2~\mathrm{nm}$ for all pairwise distances, while we scaled all lipid features using the minimal and maximal values taken along the transition.
The assembled and disassembled states are defined as configurations with $\geq 130$ interhelical contacts and with helix-helix center-of-mass distances $d_{\mathrm{CoM}} >= 3 ~\mathrm{nm}$, respectively.

The neural network used to fit the committor is implemented using keras \cite{chollet2015Keras} and consists of an initial 3-layer pyramidal part in which the number of units decreases from the 36 inputs to 6 in the last layer using a constant factor of $(6/36)^{1/2}$
followed by 6 residual units \cite{he2015DeepResidualLearning,he2016IdentityMappingsDeep}, each with 4 layers and 6 neurons per layer. A dropout of $0.01$ is applied to the inputs and the network is trained using the Adam gradient descent protocol with a learning rate of $lr=0.0001$ \cite{kingma2017AdamMethodStochastic}.

To investigate the assembly mechanism of Mga2, we distributed our reinforcement learning AI on multiple nodes of a high performance computer cluster. A single AI guided 500 independent TPS chains, each of which ran on a single computing node.  The 500 TPS simulations were initialized with random initial TPs. The neural network used to select the initial shooting points was trained on preliminary shooting attempts (8044 independent shots from 1160 different points). After two rounds (two steps in each of the 500 independent TPS chains), we updated the committor model by training on all new data. We retrained again after the sixth round. No further training was required, as indicated by consistent numbers of expected and observed counts of TPs. We performed another 14 rounds for all 500 TPS chains to harvest TPs. Shooting point selection, TPS setup and neural network training were fully automated in python code using MDAnalysis \cite{michaud-agrawal2011MDAnalysisToolkitAnalysis,gowers2016MDAnalysisPythonPackage}, numpy \cite{harris2020ArrayProgrammingNumPy} and our custom python package.

The input importance analysis revealed the total number of contacts $n_{\mathrm{contacts}}$ as the single most important input (SI Fig. \ref{fig:SI_mga2_hipr_plus}). However, no expression generated by symbolic regression as a function of $n_{\mathrm{contacts}}$ alone was accurate in reproducing the committor. It is likely that $n_{\mathrm{contacts}}$ is used by the trained network only as a binary switch to distinguish the two different regimes---close to the bound or to the unbound states. We therefore restricted the input importance analysis to training points close to the unbound state. The results reveals that the network uses various interhelical contacts that approximately retrace a helical pattern (SI Fig. \ref{fig:SI_mga2_hipr_plus}).
We performed symbolic regression on all possible combinations made by one, two, or three of the seven most important input coordinates (SI Table \ref{tab:MGA2_2d_symregs}). The best expressions in terms of the loss were selected using validation committor data that had not been used during the optimization. This validation set consists of committor data for 516 configurations with 30 trial shots each and 32 configurations with 10 trial shots.

To asses the variability in the observed reaction mechanisms, we performed a hierarchical clustering of all TPs projected into the plane defined by the contacts 9 and 22, which enter the most accurate parametrization generated by symbolic regression. We then used dynamic time warping \cite{meert2020WannesmDtaidistanceV2} to calculate the pairwise similarity between all TPs for the clustering, which we performed using the scipy clustering module \cite{scipy1.0contributors2020SciPyFundamentalAlgorithms,mullner2011ModernHierarchicalAgglomerative}.
To reflect the reactive flux \cite{e2006TheoryTransitionPaths}, the path density plots (Fig. \ref{fig:fig3}f, g) are histogrammed according to the number of paths, not the number of configurations. If a cell is visited multiple times during a path the contribution to the total histogram in this cell is still only one.

\section*{Concluding remarks: Beyond molecular self-organization}

AI-driven trajectory sampling is general 
and can immediately be adapted to sample many-body dynamics with a notion of ``likely fate'' similar to the committor. This fundamental concept of statistical mechanics extends from the game of chess \cite{krivov2011OptimalDimensionalityReductionChessDiffusion} over protein folding \cite{chung2015StructuralOriginSlow,best2005ReactionCoordinatesRates}
to climate modelling \cite{lucente2019MachineLearningCommittor}. The simulation engine---molecular dynamics in our case---is treated like a black box and can be replaced by other dynamic processes, reversible or not. Both the statistical model defining the loss function and the machine learning technology can be tailored for specific problems. More sophisticated models will be able to learn more from less data or incorporate experimental constraints. Simpler regression schemes \cite{peters2006ObtainingReactionCoordinates} can replace neural networks \cite{ma2005AutomaticMethodIdentifying} when the cost of sampling trajectories severely limits the volume of training data.

AI-driven mechanism discovery readily integrates advances in machine learning applied to force fields \cite{behler2007GeneralizedNeuralNetworkRepresentation,noe2020MachineLearningMolecular}, sampling \cite{noe2019BoltzmannGeneratorsSampling,rogal2019NeuralNetworkBasedPathCollective,sidky2020MachineLearningCollective}, and molecular representation \cite{behler2007GeneralizedNeuralNetworkRepresentation,noe2020MachineLearningMolecular,bartok2017MachineLearningUnifies}.
Increasing computational power and advances in symbolic AI will enable algorithms to distill ever more accurate mathematical descriptions of the complex processes hidden in high-dimensional data  \cite{udrescu2020AIFeynmanPhysicsInspired}.
As illustrated here, autonomous AI-driven sampling and model validation combined with symbolic AI can support the scientific discovery process.

\begin{acknowledgments}
The authors thank Prof. Christoph Dellago for stimulating discussions, Dr. Florian E. Blanc for useful comments,
and the openpathsampling community, in particular Dr. David Swenson, for discussions and technical support. H.J., R.C, and G.H. acknowledge the support of the Max Planck Society. R.C. acknowledges the support of the Frankfurt Institute for Advanced Studies. R.C. and G.H. acknowledge support by the LOEWE CMMS program of the state of Hesse. A.A. and P.G.B. acknowledge support of CSER program of the Netherlands Organization for Scientific Research (NWO) and of Shell Global Solutions International B.V. 
\end{acknowledgments}

\bibliography{biblio}

\begin{thebibliography}{10}
\expandafter\ifx\csname url\endcsname\relax
  \def\url#1{\texttt{#1}}\fi
\expandafter\ifx\csname urlprefix\endcsname\relax\def\urlprefix{URL }\fi
\providecommand{\bibinfo}[2]{#2}
\providecommand{\eprint}[2][]{\url{#2}}

\bibitem{pena-francesch2020BiosyntheticSelfhealingMaterials}
\bibinfo{author}{{Pena-Francesch}, A.}, \bibinfo{author}{Jung, H.},
  \bibinfo{author}{Demirel, M.~C.} \& \bibinfo{author}{Sitti, M.}
\newblock \bibinfo{title}{Biosynthetic self-healing materials for soft
  machines}.
\newblock \emph{\bibinfo{journal}{Nat. Mater.}} \textbf{\bibinfo{volume}{19}},
  \bibinfo{pages}{1230--1235} (\bibinfo{year}{2020}).

\bibitem{vandriessche2018MolecularNucleationMechanisms}
\bibinfo{author}{Van~Driessche, A. E.~S.} \emph{et~al.}
\newblock \bibinfo{title}{Molecular nucleation mechanisms and control
  strategies for crystal polymorph selection}.
\newblock \emph{\bibinfo{journal}{Nature}} \textbf{\bibinfo{volume}{556}},
  \bibinfo{pages}{89--94} (\bibinfo{year}{2018}).

\bibitem{chung2015StructuralOriginSlow}
\bibinfo{author}{Chung, H.~S.}, \bibinfo{author}{{Piana-Agostinetti}, S.},
  \bibinfo{author}{Shaw, D.~E.} \& \bibinfo{author}{Eaton, W.~A.}
\newblock \bibinfo{title}{Structural origin of slow diffusion in protein
  folding}.
\newblock \emph{\bibinfo{journal}{Science}} \textbf{\bibinfo{volume}{349}},
  \bibinfo{pages}{1504--1510} (\bibinfo{year}{2015}).

\bibitem{dellago1998EfficientTransitionPathSampling}
\bibinfo{author}{Dellago, C.}, \bibinfo{author}{Bolhuis, P.~G.} \&
  \bibinfo{author}{Chandler, D.}
\newblock \bibinfo{title}{Efficient transition path sampling: {{Application}}
  to {{Lennard}}-{{Jones}} cluster rearrangements}.
\newblock \emph{\bibinfo{journal}{J. Chem. Phys.}}
  \textbf{\bibinfo{volume}{108}}, \bibinfo{pages}{9236--9245}
  (\bibinfo{year}{1998}).

\bibitem{peters2006ObtainingReactionCoordinates}
\bibinfo{author}{Peters, B.} \& \bibinfo{author}{Trout, B.~L.}
\newblock \bibinfo{title}{Obtaining reaction coordinates by likelihood
  maximization}.
\newblock \emph{\bibinfo{journal}{J. Chem. Phys.}}
  \textbf{\bibinfo{volume}{125}}, \bibinfo{pages}{054108}
  (\bibinfo{year}{2006}).

\bibitem{berezhkovskii2013DiffusionSplittingCommitment}
\bibinfo{author}{Berezhkovskii, A.~M.} \& \bibinfo{author}{Szabo, A.}
\newblock \bibinfo{title}{Diffusion along the {{Splitting}}/{{Commitment
  Probability Reaction Coordinate}}}.
\newblock \emph{\bibinfo{journal}{J. Phys. Chem. B}}
  \textbf{\bibinfo{volume}{117}}, \bibinfo{pages}{13115--13119}
  (\bibinfo{year}{2013}).

\bibitem{e2006TheoryTransitionPaths}
\bibinfo{author}{E, W.} \& \bibinfo{author}{{Vanden-Eijnden}, E.}
\newblock \bibinfo{title}{Towards a {{Theory}} of {{Transition Paths}}}.
\newblock \emph{\bibinfo{journal}{J. Stat. Phys.}}
  \textbf{\bibinfo{volume}{123}}, \bibinfo{pages}{503} (\bibinfo{year}{2006}).

\bibitem{bolhuis2000ReactionCoordinatesBiomolecularIsomerization}
\bibinfo{author}{Bolhuis, P.~G.}, \bibinfo{author}{Dellago, C.} \&
  \bibinfo{author}{Chandler, D.}
\newblock \bibinfo{title}{Reaction coordinates of biomolecular isomerization}.
\newblock \emph{\bibinfo{journal}{Proc. Natl. Acad. Sci. USA}}
  \textbf{\bibinfo{volume}{97}}, \bibinfo{pages}{5877--5882}
  (\bibinfo{year}{2000}).

\bibitem{best2005ReactionCoordinatesRates}
\bibinfo{author}{Best, R.~B.} \& \bibinfo{author}{Hummer, G.}
\newblock \bibinfo{title}{Reaction coordinates and rates from transition
  paths}.
\newblock \emph{\bibinfo{journal}{Proc. Natl. Acad. Sci. USA}}
  \textbf{\bibinfo{volume}{102}}, \bibinfo{pages}{6732--6737}
  (\bibinfo{year}{2005}).

\bibitem{hummer2004TransitionPathsTransitionStates}
\bibinfo{author}{Hummer, G.}
\newblock \bibinfo{title}{From transition paths to transition states and rate
  coefficients}.
\newblock \emph{\bibinfo{journal}{J. Chem. Phys.}}
  \textbf{\bibinfo{volume}{120}}, \bibinfo{pages}{516--523}
  (\bibinfo{year}{2004}).

\bibitem{mnih2015HumanlevelControlDeep}
\bibinfo{author}{Mnih, V.} \emph{et~al.}
\newblock \bibinfo{title}{Human-level control through deep reinforcement
  learning}.
\newblock \emph{\bibinfo{journal}{Nature}} \textbf{\bibinfo{volume}{518}},
  \bibinfo{pages}{529--533} (\bibinfo{year}{2015}).

\bibitem{silver2017MasteringGameGo}
\bibinfo{author}{Silver, D.} \emph{et~al.}
\newblock \bibinfo{title}{Mastering the game of {{Go}} without human
  knowledge}.
\newblock \emph{\bibinfo{journal}{Nature}} \textbf{\bibinfo{volume}{550}},
  \bibinfo{pages}{354--359} (\bibinfo{year}{2017}).

\bibitem{ma2005AutomaticMethodIdentifying}
\bibinfo{author}{Ma, A.} \& \bibinfo{author}{Dinner, A.~R.}
\newblock \bibinfo{title}{Automatic {{Method}} for {{Identifying Reaction
  Coordinates}} in {{Complex Systems}}}.
\newblock \emph{\bibinfo{journal}{J. Phys. Chem. B}}
  \textbf{\bibinfo{volume}{109}}, \bibinfo{pages}{6769--6779}
  (\bibinfo{year}{2005}).

\bibitem{vanden-eijnden2008AssumptionsUnderlyingMilestoning}
\bibinfo{author}{{Vanden-Eijnden}, E.}, \bibinfo{author}{Venturoli, M.},
  \bibinfo{author}{Ciccotti, G.} \& \bibinfo{author}{Elber, R.}
\newblock \bibinfo{title}{On the assumptions underlying milestoning}.
\newblock \emph{\bibinfo{journal}{J. Chem. Phys.}}
  \textbf{\bibinfo{volume}{129}}, \bibinfo{pages}{174102}
  (\bibinfo{year}{2008}).

\bibitem{schmidt2009DistillingFreeFormNatural}
\bibinfo{author}{Schmidt, M.} \& \bibinfo{author}{Lipson, H.}
\newblock \bibinfo{title}{Distilling {{Free}}-{{Form Natural Laws}} from
  {{Experimental Data}}}.
\newblock \emph{\bibinfo{journal}{Science}} \textbf{\bibinfo{volume}{324}},
  \bibinfo{pages}{81--85} (\bibinfo{year}{2009}).

\bibitem{ballard2012MechanismIonicDissociation}
\bibinfo{author}{Ballard, A.~J.} \& \bibinfo{author}{Dellago, C.}
\newblock \bibinfo{title}{Toward the {{Mechanism}} of {{Ionic Dissociation}} in
  {{Water}}}.
\newblock \emph{\bibinfo{journal}{J. Phys. Chem. B}}
  \textbf{\bibinfo{volume}{116}}, \bibinfo{pages}{13490--13497}
  (\bibinfo{year}{2012}).

\bibitem{behler2007GeneralizedNeuralNetworkRepresentation}
\bibinfo{author}{Behler, J.} \& \bibinfo{author}{Parrinello, M.}
\newblock \bibinfo{title}{Generalized {{Neural}}-{{Network Representation}} of
  {{High}}-{{Dimensional Potential}}-{{Energy Surfaces}}}.
\newblock \emph{\bibinfo{journal}{Phys. Rev. Lett.}}
  \textbf{\bibinfo{volume}{98}}, \bibinfo{pages}{146401}
  (\bibinfo{year}{2007}).

\bibitem{walsh2009SimulationsSpontaneousMethaneHydrateNucleation}
\bibinfo{author}{Walsh, M.~R.}, \bibinfo{author}{Koh, C.~A.},
  \bibinfo{author}{Sloan, E.~D.}, \bibinfo{author}{Sum, A.~K.} \&
  \bibinfo{author}{Wu, D.~T.}
\newblock \bibinfo{title}{Microsecond {{Simulations}} of {{Spontaneous Methane
  Hydrate Nucleation}} and {{Growth}}}.
\newblock \emph{\bibinfo{journal}{Science}} \textbf{\bibinfo{volume}{326}},
  \bibinfo{pages}{1095--1098} (\bibinfo{year}{2009}).

\bibitem{arjun2019UnbiasedAtomisticInsight}
\bibinfo{author}{Arjun}, \bibinfo{author}{Berendsen, T.~A.} \&
  \bibinfo{author}{Bolhuis, P.~G.}
\newblock \bibinfo{title}{Unbiased atomistic insight in the competing
  nucleation mechanisms of methane hydrates}.
\newblock \emph{\bibinfo{journal}{Proc. Natl. Acad. Sci. USA}}
  \textbf{\bibinfo{volume}{116}}, \bibinfo{pages}{19305--19310}
  (\bibinfo{year}{2019}).

\bibitem{jacobson2010AmorphousPrecursorsNucleation}
\bibinfo{author}{Jacobson, L.~C.}, \bibinfo{author}{Hujo, W.} \&
  \bibinfo{author}{Molinero, V.}
\newblock \bibinfo{title}{Amorphous {{Precursors}} in the {{Nucleation}} of
  {{Clathrate Hydrates}}}.
\newblock \emph{\bibinfo{journal}{J. Am. Chem. Soc.}}
  \textbf{\bibinfo{volume}{132}}, \bibinfo{pages}{11806--11811}
  (\bibinfo{year}{2010}).

\bibitem{covino2016EukaryoticSensorMembraneSaturation}
\bibinfo{author}{Covino, R.} \emph{et~al.}
\newblock \bibinfo{title}{A {{Eukaryotic Sensor}} for {{Membrane Lipid
  Saturation}}}.
\newblock \emph{\bibinfo{journal}{Mol. Cell}} \textbf{\bibinfo{volume}{63}},
  \bibinfo{pages}{49--59} (\bibinfo{year}{2016}).

\bibitem{chiavazzo2017IntrinsicMapDynamics}
\bibinfo{author}{Chiavazzo, E.} \emph{et~al.}
\newblock \bibinfo{title}{Intrinsic map dynamics exploration for uncharted
  effective free-energy landscapes}.
\newblock \emph{\bibinfo{journal}{Proc. Natl. Acad. Sci. USA}}
  \textbf{\bibinfo{volume}{114}}, \bibinfo{pages}{E5494--E5503}
  (\bibinfo{year}{2017}).

\bibitem{bolhuis2002TRANSITIONPATHSAMPLING}
\bibinfo{author}{Bolhuis, P.~G.}, \bibinfo{author}{Chandler, D.},
  \bibinfo{author}{Dellago, C.} \& \bibinfo{author}{Geissler, P.~L.}
\newblock \bibinfo{title}{{{TRANSITION PATH SAMPLING}}: {{Throwing Ropes Over
  Rough Mountain Passes}}, in the {{Dark}}}.
\newblock \emph{\bibinfo{journal}{Annu. Rev. Phys. Chem.}}
  \textbf{\bibinfo{volume}{53}}, \bibinfo{pages}{291--318}
  (\bibinfo{year}{2002}).

\bibitem{jung2017TransitionPathSampling}
\bibinfo{author}{Jung, H.}, \bibinfo{author}{Okazaki, K.-i.} \&
  \bibinfo{author}{Hummer, G.}
\newblock \bibinfo{title}{Transition path sampling of rare events by shooting
  from the top}.
\newblock \emph{\bibinfo{journal}{J. Chem. Phys.}}
  \textbf{\bibinfo{volume}{147}}, \bibinfo{pages}{152716}
  (\bibinfo{year}{2017}).

\bibitem{kemp2007DeterminingRelativeInputImportance}
\bibinfo{author}{Kemp, S.~J.}, \bibinfo{author}{Zaradic, P.} \&
  \bibinfo{author}{Hansen, F.}
\newblock \bibinfo{title}{An approach for determining relative input parameter
  importance and significance in artificial neural networks}.
\newblock \emph{\bibinfo{journal}{Ecol. Model.}}
  \textbf{\bibinfo{volume}{204}}, \bibinfo{pages}{326--334}
  (\bibinfo{year}{2007}).

\bibitem{izzo2020DcgpDifferentiableCartesian}
\bibinfo{author}{Izzo, D.} \& \bibinfo{author}{Biscani, F.}
\newblock \bibinfo{title}{Dcgp: {{Differentiable Cartesian Genetic
  Programming}} made easy.}
\newblock \emph{\bibinfo{journal}{J. Open Source Softw.}}
  \textbf{\bibinfo{volume}{5}}, \bibinfo{pages}{2290} (\bibinfo{year}{2020}).

\bibitem{joung2008DeterminationAlkaliHalide}
\bibinfo{author}{Joung, I.~S.} \& \bibinfo{author}{Cheatham, T.~E.}
\newblock \bibinfo{title}{Determination of {{Alkali}} and {{Halide Monovalent
  Ion Parameters}} for {{Use}} in {{Explicitly Solvated Biomolecular
  Simulations}}}.
\newblock \emph{\bibinfo{journal}{J. Phys. Chem. B}}
  \textbf{\bibinfo{volume}{112}}, \bibinfo{pages}{9020--9041}
  (\bibinfo{year}{2008}).

\bibitem{jorgensen1983ComparisonSimplePotentialWater}
\bibinfo{author}{Jorgensen, W.~L.}, \bibinfo{author}{Chandrasekhar, J.},
  \bibinfo{author}{Madura, J.~D.}, \bibinfo{author}{Impey, R.~W.} \&
  \bibinfo{author}{Klein, M.~L.}
\newblock \bibinfo{title}{Comparison of simple potential functions for
  simulating liquid water}.
\newblock \emph{\bibinfo{journal}{J. Chem. Phys.}}
  \textbf{\bibinfo{volume}{79}}, \bibinfo{pages}{926--935}
  (\bibinfo{year}{1983}).

\bibitem{eastman2017OpenMM7}
\bibinfo{author}{Eastman, P.} \emph{et~al.}
\newblock \bibinfo{title}{{{OpenMM}} 7: {{Rapid}} development of high
  performance algorithms for molecular dynamics}.
\newblock \emph{\bibinfo{journal}{PLOS Comput. Biol.}}
  \textbf{\bibinfo{volume}{13}}, \bibinfo{pages}{e1005659}
  (\bibinfo{year}{2017}).

\bibitem{sivak2014TimeStepRescaling}
\bibinfo{author}{Sivak, D.~A.}, \bibinfo{author}{Chodera, J.~D.} \&
  \bibinfo{author}{Crooks, G.~E.}
\newblock \bibinfo{title}{Time {{Step Rescaling Recovers Continuous}}-{{Time
  Dynamical Properties}} for {{Discrete}}-{{Time Langevin Integration}} of
  {{Nonequilibrium Systems}}}.
\newblock \emph{\bibinfo{journal}{J. Phys. Chem. B}}
  \textbf{\bibinfo{volume}{118}}, \bibinfo{pages}{6466--6474}
  (\bibinfo{year}{2014}).

\bibitem{essmann1995SmoothParticleMeshEwald}
\bibinfo{author}{Essmann, U.} \emph{et~al.}
\newblock \bibinfo{title}{A smooth particle mesh {{Ewald}} method}.
\newblock \emph{\bibinfo{journal}{J. Chem. Phys.}}
  \textbf{\bibinfo{volume}{103}}, \bibinfo{pages}{8577--8593}
  (\bibinfo{year}{1995}).

\bibitem{behler2011AtomcenteredSymmetryFunctions}
\bibinfo{author}{Behler, J.}
\newblock \bibinfo{title}{Atom-centered symmetry functions for constructing
  high-dimensional neural network potentials}.
\newblock \emph{\bibinfo{journal}{J. Chem. Phys.}}
  \textbf{\bibinfo{volume}{134}}, \bibinfo{pages}{074106}
  (\bibinfo{year}{2011}).

\bibitem{behler2014RepresentingPotentialEnergy}
\bibinfo{author}{Behler, J.}
\newblock \bibinfo{title}{Representing potential energy surfaces by
  high-dimensional neural network potentials}.
\newblock \emph{\bibinfo{journal}{J. Phys.: Condens. Matter}}
  \textbf{\bibinfo{volume}{26}}, \bibinfo{pages}{183001}
  (\bibinfo{year}{2014}).

\bibitem{geiger2013NeuralNetworksStructureDetection}
\bibinfo{author}{Geiger, P.} \& \bibinfo{author}{Dellago, C.}
\newblock \bibinfo{title}{Neural networks for local structure detection in
  polymorphic systems}.
\newblock \emph{\bibinfo{journal}{J. Chem. Phys.}}
  \textbf{\bibinfo{volume}{139}}, \bibinfo{pages}{164105}
  (\bibinfo{year}{2013}).

\bibitem{he2015DeepResidualLearning}
\bibinfo{author}{He, K.}, \bibinfo{author}{Zhang, X.}, \bibinfo{author}{Ren,
  S.} \& \bibinfo{author}{Sun, J.}
\newblock \bibinfo{title}{Deep {{Residual Learning}} for {{Image
  Recognition}}}.
\newblock \emph{\bibinfo{journal}{arXiv:1512.03385 [cs]}}
  (\bibinfo{year}{2015}).
\newblock \eprint{1512.03385}.

\bibitem{he2016IdentityMappingsDeep}
\bibinfo{author}{He, K.}, \bibinfo{author}{Zhang, X.}, \bibinfo{author}{Ren,
  S.} \& \bibinfo{author}{Sun, J.}
\newblock \bibinfo{title}{Identity {{Mappings}} in {{Deep Residual Networks}}}.
\newblock \emph{\bibinfo{journal}{arXiv:1603.05027 [cs]}}
  (\bibinfo{year}{2016}).
\newblock \eprint{1603.05027}.

\bibitem{kingma2017AdamMethodStochastic}
\bibinfo{author}{Kingma, D.~P.} \& \bibinfo{author}{Ba, J.}
\newblock \bibinfo{title}{Adam: {{A Method}} for {{Stochastic Optimization}}}.
\newblock \emph{\bibinfo{journal}{arXiv:1412.6980 [cs]}}
  (\bibinfo{year}{2017}).
\newblock \eprint{1412.6980}.

\bibitem{chollet2015Keras}
\bibinfo{author}{Chollet, F.}
\newblock \bibinfo{title}{Keras} (\bibinfo{year}{2015}).

\bibitem{swenson2019OpenPathSampling_1Basics}
\bibinfo{author}{Swenson, D. W.~H.}, \bibinfo{author}{Prinz, J.-H.},
  \bibinfo{author}{Noe, F.}, \bibinfo{author}{Chodera, J.~D.} \&
  \bibinfo{author}{Bolhuis, P.~G.}
\newblock \bibinfo{title}{{{OpenPathSampling}}: {{A Python Framework}} for
  {{Path Sampling Simulations}}. 1. {{Basics}}}.
\newblock \emph{\bibinfo{journal}{J. Chem. Theory Comput.}}
  \textbf{\bibinfo{volume}{15}}, \bibinfo{pages}{813--836}
  (\bibinfo{year}{2019}).

\bibitem{swenson2019OpenPathSampling_2Customizing}
\bibinfo{author}{Swenson, D. W.~H.}, \bibinfo{author}{Prinz, J.-H.},
  \bibinfo{author}{Noe, F.}, \bibinfo{author}{Chodera, J.~D.} \&
  \bibinfo{author}{Bolhuis, P.~G.}
\newblock \bibinfo{title}{{{OpenPathSampling}}: {{A Python Framework}} for
  {{Path Sampling Simulations}}. 2. {{Building}} and {{Customizing Path
  Ensembles}} and {{Sample Schemes}}}.
\newblock \emph{\bibinfo{journal}{J. Chem. Theory Comput.}}
  \textbf{\bibinfo{volume}{15}}, \bibinfo{pages}{837--856}
  (\bibinfo{year}{2019}).

\bibitem{tip4pice}
\bibinfo{author}{Abascal, J. L.~F.}, \bibinfo{author}{Sanz, E.},
  \bibinfo{author}{Garc{\'i}a~Fern{\'a}ndez, R.} \& \bibinfo{author}{Vega, C.}
\newblock \bibinfo{title}{A potential model for the study of ices and amorphous
  water: {{TIP4P}}/{{Ice}}}.
\newblock \emph{\bibinfo{journal}{J. Chem. Phys.}}
  \textbf{\bibinfo{volume}{122}}, \bibinfo{pages}{234511}
  (\bibinfo{year}{2005}).

\bibitem{conde2010determining}
\bibinfo{author}{Conde, M.~M.} \& \bibinfo{author}{Vega, C.}
\newblock \bibinfo{title}{Determining the three-phase coexistence line in
  methane hydrates using computer simulations}.
\newblock \emph{\bibinfo{journal}{J. Chem. Phys.}}
  \textbf{\bibinfo{volume}{133}}, \bibinfo{pages}{064507}
  (\bibinfo{year}{2010}).

\bibitem{hess1997LINCSLinearConstraint}
\bibinfo{author}{Hess, B.}, \bibinfo{author}{Bekker, H.},
  \bibinfo{author}{Berendsen, H. J.~C.} \& \bibinfo{author}{Fraaije, J. G.
  E.~M.}
\newblock \bibinfo{title}{{{LINCS}}: {{A}} linear constraint solver for
  molecular simulations}.
\newblock \emph{\bibinfo{journal}{J. Comput. Chem.}}
  \textbf{\bibinfo{volume}{18}}, \bibinfo{pages}{1463--1472}
  (\bibinfo{year}{1997}).

\bibitem{marrink2004CoarseGrainedModel}
\bibinfo{author}{Marrink, S.~J.}, \bibinfo{author}{{de Vries}, A.~H.} \&
  \bibinfo{author}{Mark, A.~E.}
\newblock \bibinfo{title}{Coarse {{Grained Model}} for {{Semiquantitative Lipid
  Simulations}}}.
\newblock \emph{\bibinfo{journal}{J. Phys. Chem. B}}
  \textbf{\bibinfo{volume}{108}}, \bibinfo{pages}{750--760}
  (\bibinfo{year}{2004}).

\bibitem{marrink2007MARTINIForceField}
\bibinfo{author}{Marrink, S.~J.}, \bibinfo{author}{Risselada, H.~J.},
  \bibinfo{author}{Yefimov, S.}, \bibinfo{author}{Tieleman, D.~P.} \&
  \bibinfo{author}{{de Vries}, A.~H.}
\newblock \bibinfo{title}{The {{MARTINI Force Field}}:\, {{Coarse Grained
  Model}} for {{Biomolecular Simulations}}}.
\newblock \emph{\bibinfo{journal}{J. Phys. Chem. B}}
  \textbf{\bibinfo{volume}{111}}, \bibinfo{pages}{7812--7824}
  (\bibinfo{year}{2007}).

\bibitem{monticelli2008MARTINICoarseGrainedProteins}
\bibinfo{author}{Monticelli, L.} \emph{et~al.}
\newblock \bibinfo{title}{The {{MARTINI Coarse}}-{{Grained Force Field}}:
  {{Extension}} to {{Proteins}}}.
\newblock \emph{\bibinfo{journal}{J. Chem. Theory Comput.}}
  \textbf{\bibinfo{volume}{4}}, \bibinfo{pages}{819--834}
  (\bibinfo{year}{2008}).

\bibitem{dejong2013ImprovedParametersMartini}
\bibinfo{author}{{de Jong}, D.~H.} \emph{et~al.}
\newblock \bibinfo{title}{Improved {{Parameters}} for the {{Martini
  Coarse}}-{{Grained Protein Force Field}}}.
\newblock \emph{\bibinfo{journal}{J. Chem. Theory Comput.}}
  \textbf{\bibinfo{volume}{9}}, \bibinfo{pages}{687--697}
  (\bibinfo{year}{2013}).

\bibitem{berendsen1995GROMACSMessagepassingParallel}
\bibinfo{author}{Berendsen, H.}, \bibinfo{author}{{van der Spoel}, D.} \&
  \bibinfo{author}{{van Drunen}, R.}
\newblock \bibinfo{title}{{{GROMACS}}: {{A}} message-passing parallel molecular
  dynamics implementation}.
\newblock \emph{\bibinfo{journal}{Comput. Phys. Commun.}}
  \textbf{\bibinfo{volume}{91}}, \bibinfo{pages}{43--56}
  (\bibinfo{year}{1995}).

\bibitem{hess2008GROMACS4}
\bibinfo{author}{Hess, B.}, \bibinfo{author}{Kutzner, C.},
  \bibinfo{author}{{van der Spoel}, D.} \& \bibinfo{author}{Lindahl, E.}
\newblock \bibinfo{title}{{{GROMACS}} 4: {{Algorithms}} for {{Highly
  Efficient}}, {{Load}}-{{Balanced}}, and {{Scalable Molecular Simulation}}}.
\newblock \emph{\bibinfo{journal}{J. Chem. Theory Comput.}}
  \textbf{\bibinfo{volume}{4}}, \bibinfo{pages}{435--447}
  (\bibinfo{year}{2008}).

\bibitem{pronk2013GROMACS4.5}
\bibinfo{author}{Pronk, S.} \emph{et~al.}
\newblock \bibinfo{title}{{{GROMACS}} 4.5: A high-throughput and highly
  parallel open source molecular simulation toolkit}.
\newblock \emph{\bibinfo{journal}{Bioinformatics}}
  \textbf{\bibinfo{volume}{29}}, \bibinfo{pages}{845--854}
  (\bibinfo{year}{2013}).

\bibitem{abraham2015GROMACSHighPerformance}
\bibinfo{author}{Abraham, M.~J.} \emph{et~al.}
\newblock \bibinfo{title}{{{GROMACS}}: {{High}} performance molecular
  simulations through multi-level parallelism from laptops to supercomputers}.
\newblock \emph{\bibinfo{journal}{SoftwareX}} \textbf{\bibinfo{volume}{1-2}},
  \bibinfo{pages}{19--25} (\bibinfo{year}{2015}).

\bibitem{bussi2007CanonicalSamplingVelocity}
\bibinfo{author}{Bussi, G.}, \bibinfo{author}{Donadio, D.} \&
  \bibinfo{author}{Parrinello, M.}
\newblock \bibinfo{title}{Canonical sampling through velocity rescaling}.
\newblock \emph{\bibinfo{journal}{J. Chem. Phys.}}
  \textbf{\bibinfo{volume}{126}}, \bibinfo{pages}{014101}
  (\bibinfo{year}{2007}).

\bibitem{parrinello1981PolymorphicTransitionsSingle}
\bibinfo{author}{Parrinello, M.} \& \bibinfo{author}{Rahman, A.}
\newblock \bibinfo{title}{Polymorphic transitions in single crystals: {{A}} new
  molecular dynamics method}.
\newblock \emph{\bibinfo{journal}{J. Appl. Phys.}}
  \textbf{\bibinfo{volume}{52}}, \bibinfo{pages}{7182--7190}
  (\bibinfo{year}{1981}).

\bibitem{michaud-agrawal2011MDAnalysisToolkitAnalysis}
\bibinfo{author}{{Michaud-Agrawal}, N.}, \bibinfo{author}{Denning, E.~J.},
  \bibinfo{author}{Woolf, T.~B.} \& \bibinfo{author}{Beckstein, O.}
\newblock \bibinfo{title}{{{MDAnalysis}}: {{A}} toolkit for the analysis of
  molecular dynamics simulations}.
\newblock \emph{\bibinfo{journal}{J. Comput. Chem.}}
  \textbf{\bibinfo{volume}{32}}, \bibinfo{pages}{2319--2327}
  (\bibinfo{year}{2011}).

\bibitem{gowers2016MDAnalysisPythonPackage}
\bibinfo{author}{Gowers, R.} \emph{et~al.}
\newblock \bibinfo{title}{{{MDAnalysis}}: {{A Python Package}} for the {{Rapid
  Analysis}} of {{Molecular Dynamics Simulations}}}.
\newblock In \emph{\bibinfo{booktitle}{Python in {{Science Conference}}}},
  \bibinfo{pages}{98--105} (\bibinfo{address}{{Austin, Texas}},
  \bibinfo{year}{2016}).

\bibitem{harris2020ArrayProgrammingNumPy}
\bibinfo{author}{Harris, C.~R.} \emph{et~al.}
\newblock \bibinfo{title}{Array programming with {{NumPy}}}.
\newblock \emph{\bibinfo{journal}{Nature}} \textbf{\bibinfo{volume}{585}},
  \bibinfo{pages}{357--362} (\bibinfo{year}{2020}).

\bibitem{meert2020WannesmDtaidistanceV2}
\bibinfo{author}{Meert, W.}, \bibinfo{author}{Hendrickx, K.} \&
  \bibinfo{author}{Van~Craenendonck, T.}
\newblock \bibinfo{title}{Wannesm/dtaidistance v2.0.0}.
\newblock \bibinfo{howpublished}{http://doi.org/10.5281/zenodo.3981067}
  (\bibinfo{year}{2020}).

\bibitem{scipy1.0contributors2020SciPyFundamentalAlgorithms}
\bibinfo{author}{{SciPy 1.0 Contributors}} \emph{et~al.}
\newblock \bibinfo{title}{{{SciPy}} 1.0: Fundamental algorithms for scientific
  computing in {{Python}}}.
\newblock \emph{\bibinfo{journal}{Nat. Methods}} \textbf{\bibinfo{volume}{17}},
  \bibinfo{pages}{261--272} (\bibinfo{year}{2020}).

\bibitem{mullner2011ModernHierarchicalAgglomerative}
\bibinfo{author}{M{\"u}llner, D.}
\newblock \bibinfo{title}{Modern hierarchical, agglomerative clustering
  algorithms}.
\newblock \emph{\bibinfo{journal}{arXiv:1109.2378 [cs, stat]}}
  (\bibinfo{year}{2011}).
\newblock \eprint{1109.2378}.

\bibitem{krivov2011OptimalDimensionalityReductionChessDiffusion}
\bibinfo{author}{Krivov, S.~V.}
\newblock \bibinfo{title}{Optimal dimensionality reduction of complex dynamics:
  {{The}} chess game as diffusion on a free-energy landscape}.
\newblock \emph{\bibinfo{journal}{Phys. Rev. E}} \textbf{\bibinfo{volume}{84}},
  \bibinfo{pages}{011135} (\bibinfo{year}{2011}).

\bibitem{lucente2019MachineLearningCommittor}
\bibinfo{author}{Lucente, D.}, \bibinfo{author}{Duffner, S.},
  \bibinfo{author}{Herbert, C.}, \bibinfo{author}{Rolland, J.} \&
  \bibinfo{author}{Bouchet, F.}
\newblock \bibinfo{title}{Machine learning of committor functions for
  predicting high impact climate events}.
\newblock \emph{\bibinfo{journal}{arXiv:1910.11736}}  (\bibinfo{year}{2019}).
\newblock \eprint{1910.11736}.

\bibitem{noe2020MachineLearningMolecular}
\bibinfo{author}{No{\'e}, F.}, \bibinfo{author}{Tkatchenko, A.},
  \bibinfo{author}{M{\"u}ller, K.-R.} \& \bibinfo{author}{Clementi, C.}
\newblock \bibinfo{title}{Machine {{Learning}} for {{Molecular Simulation}}}.
\newblock \emph{\bibinfo{journal}{Annu. Rev. Phys. Chem.}}
  \textbf{\bibinfo{volume}{71}}, \bibinfo{pages}{361--390}
  (\bibinfo{year}{2020}).

\bibitem{noe2019BoltzmannGeneratorsSampling}
\bibinfo{author}{No{\'e}, F.}, \bibinfo{author}{Olsson, S.},
  \bibinfo{author}{K{\"o}hler, J.} \& \bibinfo{author}{Wu, H.}
\newblock \bibinfo{title}{Boltzmann generators: {{Sampling}} equilibrium states
  of many-body systems with deep learning}.
\newblock \emph{\bibinfo{journal}{Science}} \textbf{\bibinfo{volume}{365}}
  (\bibinfo{year}{2019}).

\bibitem{rogal2019NeuralNetworkBasedPathCollective}
\bibinfo{author}{Rogal, J.}, \bibinfo{author}{Schneider, E.} \&
  \bibinfo{author}{Tuckerman, M.~E.}
\newblock \bibinfo{title}{Neural-{{Network}}-{{Based Path Collective
  Variables}} for {{Enhanced Sampling}} of {{Phase Transformations}}}.
\newblock \emph{\bibinfo{journal}{Phys. Rev. Lett.}}
  \textbf{\bibinfo{volume}{123}}, \bibinfo{pages}{245701}
  (\bibinfo{year}{2019}).

\bibitem{sidky2020MachineLearningCollective}
\bibinfo{author}{Sidky, H.}, \bibinfo{author}{Chen, W.} \&
  \bibinfo{author}{Ferguson, A.~L.}
\newblock \bibinfo{title}{Machine learning for collective variable discovery
  and enhanced sampling in biomolecular simulation}.
\newblock \emph{\bibinfo{journal}{Mol. Phys.}} \textbf{\bibinfo{volume}{118}},
  \bibinfo{pages}{e1737742} (\bibinfo{year}{2020}).

\bibitem{bartok2017MachineLearningUnifies}
\bibinfo{author}{Bart{\'o}k, A.~P.} \emph{et~al.}
\newblock \bibinfo{title}{Machine learning unifies the modeling of materials
  and molecules}.
\newblock \emph{\bibinfo{journal}{Sci. Adv.}} \textbf{\bibinfo{volume}{3}},
  \bibinfo{pages}{e1701816} (\bibinfo{year}{2017}).

\bibitem{udrescu2020AIFeynmanPhysicsInspired}
\bibinfo{author}{Udrescu, S.-M.} \& \bibinfo{author}{Tegmark, M.}
\newblock \bibinfo{title}{{{AI Feynman}}: {{A}} physics-inspired method for
  symbolic regression}.
\newblock \emph{\bibinfo{journal}{Sci. Adv.}} \textbf{\bibinfo{volume}{6}},
  \bibinfo{pages}{eaay2631} (\bibinfo{year}{2020}).

\bibitem{barnes2014two}
\bibinfo{author}{Barnes, B.~C.}, \bibinfo{author}{Beckham, G.~T.},
  \bibinfo{author}{Wu, D.~T.} \& \bibinfo{author}{Sum, A.~K.}
\newblock \bibinfo{title}{Two-component order parameter for quantifying
  clathrate hydrate nucleation and growth}.
\newblock \emph{\bibinfo{journal}{J. Chem. Phys.}}
  \textbf{\bibinfo{volume}{140}}, \bibinfo{pages}{164506}
  (\bibinfo{year}{2014}).

\bibitem{rodger1996simulations}
\bibinfo{author}{Rodger, P.~M.}, \bibinfo{author}{Forester, T.~R.} \&
  \bibinfo{author}{Smith, W.}
\newblock \bibinfo{title}{Simulations of the methane hydrate/methane gas
  interface near hydrate forming conditions conditions}.
\newblock \emph{\bibinfo{journal}{Fluid Phase Equil.}}
  \textbf{\bibinfo{volume}{116}}, \bibinfo{pages}{326--332}
  (\bibinfo{year}{1996}).

\end{thebibliography}

\newpage
\setcounter{figure}{0}
\renewcommand{\figurename}{SI Figure}
\renewcommand{\tablename}{SI Table}
\renewcommand{\thetable}{S\arabic{table}}
\renewcommand{\thefigure}{S\arabic{figure}}

\section*{Supplementary Information}
\subsection*{Normalization of symmetry functions}
\label{sec:SI_cos_integral}
\paragraph*{Type $G^2$.}
The symmetry functions of type $G^2$ count the number of solvent atoms in the probing volume, the normalization constant $\langle N[G_i^2]\rangle _{\mathrm{iso}}$ is therefore the expected number of atoms in the probing volume $V_{\mathrm{probe}}^{(2)}$,
\begin{equation*}
\langle N[G_i^2]\rangle _{\mathrm{iso}} = \rho_N V_{\mathrm{probe}}^{(2)},
\end{equation*}
where $\rho_N$ is the average number density of the probed solvent atom type. 
The exact probing volume for the $G^2$ type can be approximated as
\begin{align*}
V_{\mathrm{probe}}^{(2)} &= \int_{0}^{\infty} dr \int_{0}^{\pi} d\theta \int_{0}^{2\pi} d\phi ~r^2 \sin(\theta) \nonumber\\ &\quad\quad\times \exp(-\eta(r - r_s)^2) ~f_c(r) \nonumber\\
&\approx 8 \pi r_s^2 \sqrt{2/\eta}.
\end{align*}
for small $\eta$ and $r_{\mathrm{cut}}>r_s$.

\paragraph*{Type $G^5$.}
The functions of type $G^5$ include an additional angular term and count the number of solvent atom pairs located on opposite sides of the central solute atom. The expected number of pairs $\langle N_{\mathrm{pairs}}\rangle _{\mathrm{iso}}$ can be calculated from the expected number of atoms in the probed volume $\langle N_{\mathrm{atoms}}\rangle _{\mathrm{iso}}$ as $ \langle N_{\mathrm{atoms}}\rangle _{\mathrm{iso}} (\langle N_{\mathrm{atoms}}\rangle _{\mathrm{iso}} -1)/2$. This expression is only exact for integer values of $\langle N_{\mathrm{atoms}}\rangle _{\mathrm{iso}}$ and can even become negative if $\langle N_{\mathrm{atoms}}\rangle _{\mathrm{iso}} < 1$. We therefore used an approximation which is guaranteed to be non-negative,
\begin{equation*}
\langle N_{\mathrm{pairs}}\rangle _{\mathrm{iso}} \approx \frac{\langle N_{\mathrm{atoms}}\rangle _{\mathrm{iso}}^2}{2}.
\end{equation*}
The expected number of atoms $\langle N_{\mathrm{atoms}}\rangle _{\mathrm{iso}}$ can be calculated from the volume that is probed for a fixed solute atom and with one fixed solvent atom,
\begin{align*}
V_{\mathrm{probe}}^{(5)} &= 2^{1-\zeta} \int_{0}^{\infty} dr \int_{0}^{\pi} d\theta \int_{0}^{2\pi} d\phi ~r^2 \sin(\theta) (1 \pm \cos(\phi))^\zeta\nonumber\\ &\quad\quad\times \exp(-\eta(r - r_s)^2) ~f_c(r)\nonumber\\
&=2^{1-\zeta} V_{\mathrm{probe}}^{(2)} \frac{(2\zeta-1)!!}{\zeta!}
\end{align*}

\begin{table}
	\centering
	\caption{Symmetry functions used to describe solvent configurations in the formation of lithium-chloride ion pairs. 
	For symmetry functions of type $G^5$ we used a total of 10 different parameter combinations for each value of $r_s$. A cutoff of $r_{\mathrm{cut}} = 1 ~\mathrm{nm}$ is used in the Fermi cutoff function.}
	\label{tab:LiCl_sf_params}
	\begin{tabular}{l|c|c|c|c}
		\toprule
		& \multicolumn{4}{c}{Symmetry function type} \\
		& $G^2(r_s, \eta)$ & \multicolumn{3}{c}{$G^5(r_s, \eta, \zeta, \lambda)$}\\
		\midrule
		$r_s ~[\mathrm{nm}]$ & $\eta$ & $\eta$ & $\zeta$ & $\lambda$\\
		\midrule
		0.175 & \multirow{6}{*}{200} & \multirow{6}{*}{120} & \multirow{6}{*}{1, 2, 4, 16, 64} & \multirow{6}{*}{+1, -1} \\
		0.25 & & & & \\
		0.4 & & & & \\
		0.55 & & & & \\
		0.7 & & & & \\		
		0.85 & & & & \\
		\bottomrule
	\end{tabular}
\end{table}
\begin{table}
	\centering 
	\caption{Input relevance analysis for lithium-chloride ion-pair formation. Definition of the ten most relevant input coordinates, listed in decreasing order of importance.}
	\label{tab:LiCl_important_inputs_by_relevance}
	\begin{tabular}{lc}
		\toprule
		Index & Definition \\
		$x_{220}$ & $r_{\mathrm{LiCl}}$ \\
    	$x_{12}$ & $G^5_{\mathrm{Li}}(\eta=120.0, r_s=0.25, \zeta=16, \lambda=-1.0)$[O of HOH] \\
		$x_8$ & $G^5_{\mathrm{Li}}(\eta=120.0, r_s=0.25, \zeta=2, \lambda=-1.0)$[O of HOH] \\
		$x_{14}$ & $G^5_{\mathrm{Li}}(\eta=120.0, r_s=0.25, \zeta=64, \lambda=-1.0)$[O of HOH] \\
		$x_{55}$ & $G^2_{\mathrm{Cl}}(\eta=200.0, r_s=0.25)$[O of HOH] \\
		$x_6$ & $G^5_{\mathrm{Li}}(\eta=120.0, r_s=0.25, \zeta=1, \lambda=-1.0)$[O of HOH] \\
		$x_9$ & $G^5_{\mathrm{Li}}(\eta=120.0, r_s=0.25, \zeta=4, \lambda=1.0)$[O of HOH] \\
		$x_{173}$ & $G^5_{\mathrm{Cl}}(\eta=120.0, r_s=0.25, \zeta=2, \lambda=-1.0)$[H of HOH] \\
		$x_{178}$ & $G^5_{\mathrm{Cl}}(\eta=120.0, r_s=0.25, \zeta=64, \lambda=1.0)$[H of HOH] \\
    	$x_{110}$ & $G^2_{\mathrm{Li}}(\eta=200.0, r_s=0.25)$[H of HOH] \\
    	\bottomrule
	\end{tabular}
\end{table}

\begin{table*}
\caption{Overview of the features used to describe the methane clathrate nucleation. Features are grouped by category, the indices are used for the input to neural networks and symbolic regression.}
\label{tab:clathrate_descriptors}
    \addtolength{\leftskip} {-2cm}
    \addtolength{\rightskip}{-2cm}
\centering
\begin{tabular}{ |p{2.5cm}||p{2.5cm}|p{1cm}|p{10.5cm}| }
 \hline
 Category & Name & Index & Definition\\
 \hline
 \multirow{5}{6em}{Methanes in nucleus} & MCG & 0 & Total number of
 methanes in the largest cluster\cite{barnes2014two}\\
 & N$_{sm\_1}$ & 11  & Methanes (in MCG) with only 1 methane neighbor
within 0.9 nm\\
 & N$_{sm\_2}$ & 12 & Methanes (in MCG) with 1 or 2 methane neighbor
within 0.9 nm\\\
 & N$_{cm\_1}$ & 13 & Core methanes:  MCG minus N$_{sm\_1}$\\
 & N$_{cm\_2}$ & 14 & Core methanes:  MCG minus N$_{sm\_2}$\\
 \hline
 \multirow{5}{6em}{Waters molecules in the nucleus} & N$_{w\_2}$ & 3 & Number of waters with 2 MCG
carbons within 0.6 nm\\
 & N$_{w\_3}$ & 2 & Number of waters with 3 MCG carbons within 0.6 nm\\
 & N$_{w\_4}$ & 1 & Number of waters with 4 MCG carbons within 0.6 nm\\
 & N$_{sw\_2-3}$ & 5 & Surface water molecules 1: N$_{w\_2}-$ N$_{w\_3}$ \\
 & N$_{sw\_3-4}$ & 4 & Surface water molecules 2: N$_{w\_3}-$ N$_{w\_4}$ \\
 \hline
 \multirow{4}{6em}{Structure of the nucleus} & 5$^{12}$6$^2$ cages & 8 &
Cages with 12 planar five-rings and 2 planar six-rings\\
 & 5$^{12}$ cages & 9 & Cages with 12 planar five-rings\\
  & 5$^{12}$6$^3$ cages & 17 & Cages with 12 planar five-rings and 3 planar
six-rings\\
    & 5$^{12}$6$^4$ cages & 18 & Cages with 12 planar five-rings and 3
planar six-rings\\
     & 4$^{1}$5$^{12}$6$^2$ cages & 19 & Cages with 1 planar four-ring, 12
five-rings and 2 six-rings\\
      & 4$^{1}$5$^{12}$6$^3$ cages & 20 & Cages with 1 planar four-ring, 12
five-rings and 3 six-rings\\
       & 4$^{1}$5$^{12}$6$^4$ cages & 21 & Cages with 1 planar four-ring,
12 five-rings and 4 six-rings\\
 & Cage Ratio & 10 & 5$^{12}$6$^2$ cages divided by 5$^{12}$ cages\\
 & R$_g$ & 7 & Radius of gyration of the nucleus\\
 \hline
 Global Crystallinity & F4 & 6 & Average of 3 times the cosine of the
dihedral angle between two neighboring waters.\cite{rodger1996simulations}\\
 \hline
\end{tabular}
\end{table*}
\begin{table*}
    \caption{Number of shooting results and outcomes for all temperatures}
    \label{tab:clathrates_dataset}
    \centering
    \begin{tabular}{c | c c | c c}
        \toprule
        & \multicolumn{2}{c|}{TPS} & \multicolumn{2}{c}{Committor validation} \\
        \hline
        \multirow{2}{6em}{Temperature} & \multirow{2}{8em}{Configurations} & Shooting results & \multirow{2}{8em}{Configurations} & Shooting results \\
        & & (A | B) & & (A | B) \\
        \hline
        270 K & 661 &  357 | 304 & 35 & 289 | 258 \\
        275 K & 558 & 259 | 299 & 39 & 356 | 313 \\
        280 K & 982 & 536 | 446 & 53 & 304 | 255  \\
        285 K & 1197 & 646 | 551 & 33 & 280 | 299 \\
        \hline
        all & 3398 & 1798 | 1600 & 160 & 1229 | 1125 \\
        \bottomrule
    \end{tabular}
\end{table*}

\begin{table*}
\caption{Features used to describe MGA2 transmembrane assembly}
\label{tab:MGA2_descriptors}
\centering
\begin{tabular}{ |p{2.5cm}||p{2.5cm}|p{12cm}| } 
 \hline
 Category & Name & Definition\\
 \hline
 \makecell[lc]{Pairwise \\ interhelical \\ contacts} & $x_{0}$ - $x_{27}$ & $x_{i}(r_{i}) = \frac{1 - ( r_i / (2 ~\mathrm{nm}))^6}{1 - ( r_i / (2 ~\mathrm{nm}))^{12}}$, 
 where $r_i$ is the distance between the $i$th residue on each helix; index 0 corresponds to ASN1034, index 28 to GLN1061 \\
  \hline
 \multirow{3}{2.5cm}{Global conformation} & $x_{28} = n_{\mathrm{contacts}}$ & $n_{\mathrm{contacts}}(\mathbf{r}) = \sum\limits_{i=0}^{27} \frac{1 - (r_i - 0.07~\mathrm{nm})/(0.7 ~\mathrm{nm}))^{6}}{1 - (r_i - 0.07~\mathrm{nm})/(0.7 ~\mathrm{nm}))^{30}}$\\
 & $x_{29} = \alpha_{tilt}$ & Angle between the first principal moments of inertia of the two helices\\
 & $x_{30} = d_{CoM}$ & Center of mass distance between the two helices in the plane of the membrane\\
 \hline
 \multirow{5}{2.5cm}{Lipid collective variables} & $x_{31} = \mathrm{CV}_{lip1}$ & Number of lipid tails crossing the helix-helix interface\\
 & $x_{32} = \mathrm{CV}_{lip2}$ & Number of lipid molecules that cross the interface with both tails\\
  & $x_{33} = \mathrm{CV}_{lip3}$ & Number of lipid molecules with center of mass in the helix-helix interface\\
  & $x_{34} = \mathrm{CV}_{lip4}$ & Number of lipid molecules with the headgroup in the helix-helix interface \\
  & $x_{35} = \mathrm{CV}_{lip5}$ & Number of lipid molecules with their headgroup in the interface and the tails spread in opposite directions\\
 \hline
\end{tabular}
\end{table*}

\begin{table*}
\caption{MGA2 symbolic regression results for each possible 2 coordinate combination of the seven most relevant inputs. Combinations including $n_{\mathrm{contacts}}$ are omitted due to their low predictive power.}
\label{tab:MGA2_2d_symregs}
\centering
\begin{tabular}{ l  l }
 \toprule
 Validation loss & \makecell[c]{Expression} \\
 \hline
 $0.53004$ & $q_B(x_9, x_{22}) = - \exp(x_9 ^2)\log(x_9 - \frac{x_9}{\log(x_{22})})$\\
 $0.54033$ & $q_B(x_{1}, x_{22}) = (\exp(x_{22}) + 0.637)(-0.0557\exp(2 x_{22}) - \log(x_{22} + x_{1}))$\\
 $0.53687$ & $q_B(x_9, x_{26}) = - x_{26} - \frac{x_{26}}{0.26/x_9 - \log(x_9)} + \log(0.26/x_9) + 1.29$\\
 $0.54909$ & $q_B(x_9, x_{23}) = \left(\frac{9.07 x_9}{9.07 x_9 - 25.9} - 1.4\right)(x_{23} + 2 x_9 - 1.4) $\\
 $0.55821$ & $q_B(x_{1}, x_{20}) = 2.54 - 2.77x_{20} - 2 x_1 - 0.287\exp(x_1)$\\
 $0.55709$ & $q_B(x_1, x_{26}) = (\exp(-2.07 x_1) - x_{26})\log(21.3\exp(-x_1 (1.07 - x_1) - 0.00272 x_1))$\\
 $0.56051$ & $q_B(x_9, x_{20}) = 2.67 - \exp(x_{20}) - \frac{7.29}{-1.49 + 4.13/x_9}$\\
 $0.55964$ & $q_B(x_{1}, x_{23}) = 0.00426(-10 x_{23} - 64.4\exp(x_{23}) + 924)\exp(-x_1) - 1.43\exp(x_{23})$\\
 $0.57002$ & $q_B(x_1, x_9) = - \log\left(-0.0626 - \frac{0.816}{\log(x_9)}\right)$\\
 $0.5956$ & $q_B(x_{22}, x_{26}) = 1.24 - \frac{2.76 x_{22}^2}{x_{26}}$\\
 $0.58008$ & $q_B(x_{20}, x_{22}) = - 0.496\exp(x_{20} + x_{22})\log(x_{20} + x_{22})$\\
 $0.59576$ & $q_B(x_{22}, x_{23}) = - (0.605\exp(x_{22}) - 0.0956)\log(x_{22}\exp(x_{22})) - 0.282$\\
 $0.59914$ & $q_B(x_{20}, x_{26}) = \frac{\log(2 x_{20})}{x_{20} + x_{26} - 2.3}$\\
 $0.5975$ & $q_B(x_{20}, x_{23}) = 1.73 - \exp(x_{20}) + \frac{0.00635 x_{20}(x_{20} + 6.11)}{\log(x_{23})}$\\
 $0.61409$ & $q_B(x_{23}, x_{26}) = \exp(-5.89 x_{23} + \frac{0.0193}{x_{23}}) - \frac{8.31 x_{23}\exp(\exp(x_{26}))}{-2.72 + 78.7/x_{23}^2}$\\
 \bottomrule
\end{tabular}
\end{table*}

\begin{figure*}
    \centering
        \includegraphics[width=0.7\textwidth]{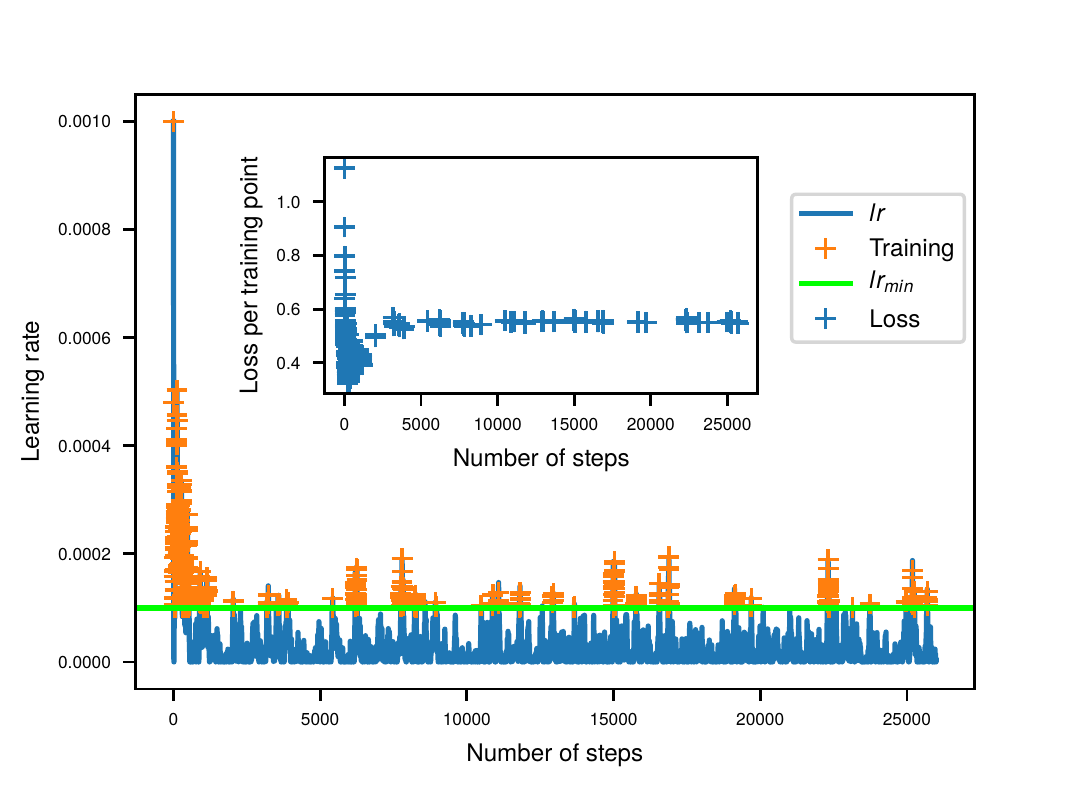}
    \caption{Training iterations for LiCl assembly. The blue line shows the learning rate calculated from the efficiency factor at every step, orange crosses show when training actually occurred. The inset shows the training loss per shooting point for every training.}
    \label{fig:SI_LiCl_training_loss}
\end{figure*}

\begin{figure*}
    \centering
        \includegraphics[width=0.7\textwidth]{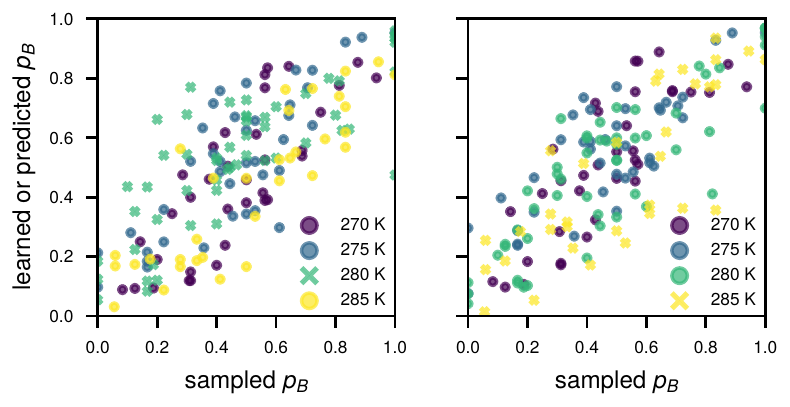}
    \caption{Interpolation and extrapolation in temperatures of methane clathrate nucleation models.
    Cross-correlation between learned committor and the committor as obtained by repeated sampling for two models which are trained on only three of the four temperatures available in the training set. (Left) Committor model trained only on $T=270$ K, $275$ K and $285$~K , i.e. leaving out $T=280$~K, to assess the model's ability to interpolate. (Right) Model trained on $T=270$ K, $275$ K and $280$~K to assess the model's ability to extrapolate.}
    \label{fig:SI_clathrates_leave_one_T_out}
\end{figure*}

\begin{figure*}
    \centering
        \includegraphics[width=0.7\textwidth]{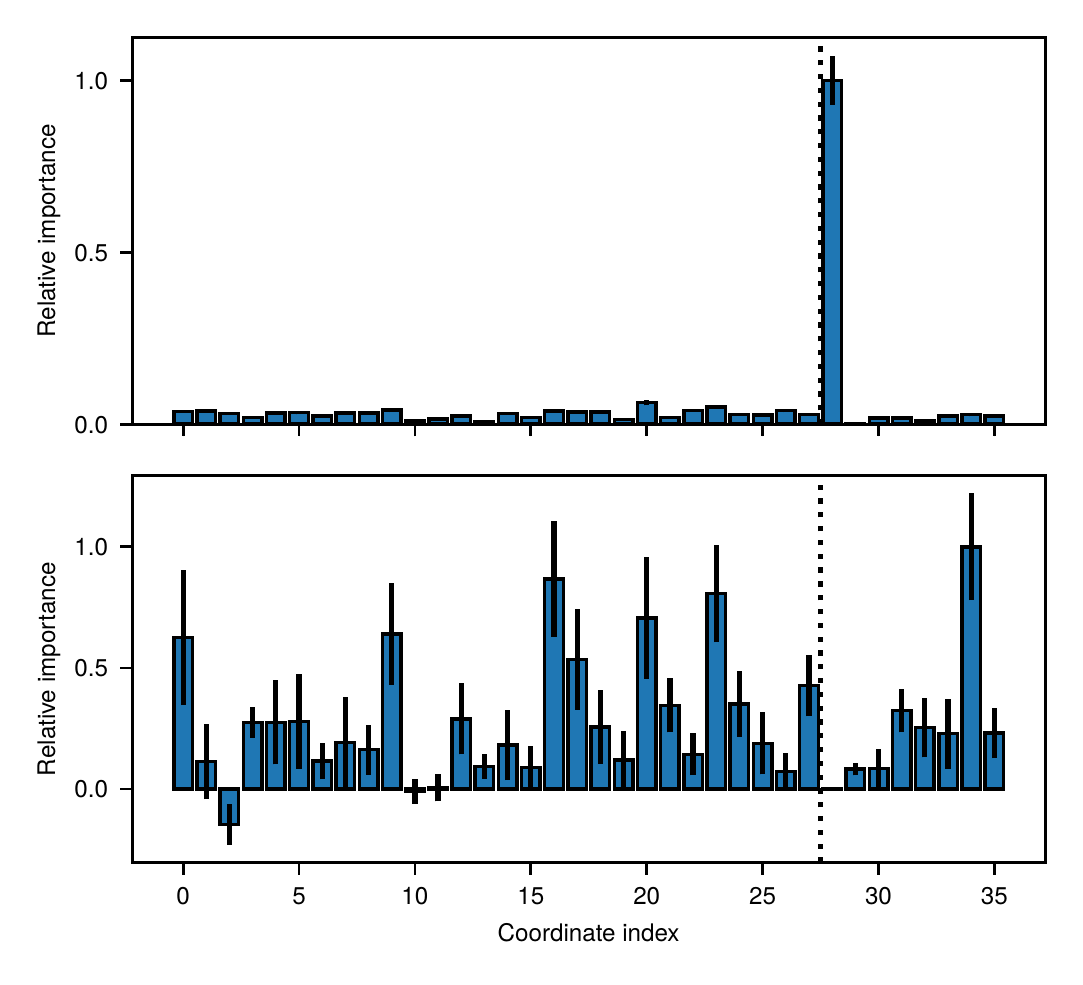}
    \caption{Input importance analyses for Mga2 transmembrane assembly, by using all training points (top panel), and a subset with $n_{\mathrm{contacts}} < 0.01$ (bottom panel), corresponding to training points close to the unbound state. The height of each bar is the average over 50 independent analyses, while the bars indicate one standard deviation. All values are normalized to the largest importance in each set.}
    \label{fig:SI_mga2_hipr_plus}
\end{figure*}

\begin{figure*}
    \centering
        \includegraphics[width=0.5\textwidth]{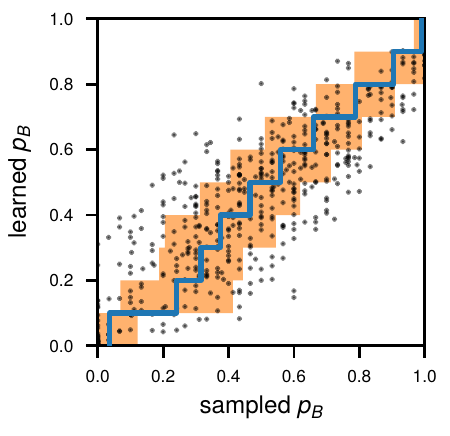}
    \caption{Committor cross correlation plot for the symbolic regression expression $q_B(x_9, x_{22})$ on untrained validation committor data. The average of the sampled committors (blue line) and their standard deviation (orange shaded) are calculated by binning along the predicted committor.}
    \label{fig:SI_mga2_best_symreg_pB_cross}
\end{figure*}

\clearpage

\end{document}